\documentclass[
showpacs,
amsmath,amssymb,
aps,
prb,
twocolumn,
]{revtex4-1}

\usepackage{bibunits}
\usepackage{graphicx}
\usepackage{dcolumn}
\usepackage{bm}
\usepackage[breaklinks,colorlinks = true,linkcolor = red,urlcolor=blue,citecolor=red]{hyperref}
\usepackage{multirow}
\usepackage{array}
\usepackage{booktabs}
\usepackage{ctable}
\usepackage{upgreek}
\usepackage{epsfig}
\usepackage{mathrsfs}
\usepackage{amssymb}
\usepackage{amsbsy}
\usepackage{color}
\usepackage{cancel}
\usepackage{pifont}
\usepackage{marginnote}
\usepackage{float}
\usepackage{verbatim}
\usepackage{subfigure}
\usepackage{bbm, dsfont}
\usepackage{lineno}

\newcommand{\bcen}{\begin{center}}
\newcommand{\ecen}{\end{center}}
\newcommand{\btab}{\begin{tabular}}
\newcommand{\etab}{\end{tabular}}
\newcommand{\bdes}{\begin{description}}
\newcommand{\edes}{\end{description}}

\newcommand{\beq}{\begin{equation}}
\newcommand{\eeq}{\end{equation}}
\newcommand{\bea}{\begin{eqnarray}}
\newcommand{\eea}{\end{eqnarray}}

\newcommand{\bary}{\begin{array}}
	\newcommand{\eary}{\end{array}}
\newcommand{\benum}{\begin{enumerate}}
	\newcommand{\eenum}{\end{enumerate}}
\newcommand{\bitem}{\begin{itemize}}
	\newcommand{\eitem}{\end{itemize}}

\newcommand{\bOne}{{\boldsymbol 1}}
\newcommand{\br} { \boldsymbol{r}}

\newcommand{\bA} { \mbox{\boldmath $A$}}

\newcommand{\eqn}[1] {eqn.~(\ref{#1})}

\newcommand{\Fig}[1]{Fig.~\ref{#1}}

\makeatletter

\newcommand{\Rmnum}[1]{\expandafter\@slowromancap\romannumeral #1@}
\makeatother

\newcommand{\titlename}{Topological and conventional phases of a three dimensional electron glass}

\begin{document}

\title{\titlename}

\author{Prateek Mukati}
\email{prateek.mukati@icts.res.in}
\affiliation{International Centre for Theoretical Sciences, Tata Institute of Fundamental Research, Bengaluru 560089, India}
\author{Adhip Agarwala}
\email{adhip.agarwala@icts.res.in}
\affiliation{International Centre for Theoretical Sciences, Tata Institute of Fundamental Research, Bengaluru 560089, India}
\author{Subhro Bhattacharjee}
\email{subhro@icts.res.in}
\affiliation{International Centre for Theoretical Sciences, Tata Institute of Fundamental Research, Bengaluru 560089, India}



\date{\today}

\begin{abstract}


We investigate a symmetry protected $Z_2$  topological electron glass -- a glassy equivalent of the $Z_2$ topological band insulator in crystalline systems-- and uncover associated quantum phase transitions in this three dimensional amorphous network of atoms. Through explicit numerical calculations of the Witten effect, we show that  the $Z_2$ glass is characterized by an anomalous electromagnetic response-- dyons with $1/2$ electronic charge. We further study, using a variety of numerical diagnostics including such electromagnetic responses, the phase transitions of the $Z_2$ glass in to a metallic and/or a trivial insulating phase. We find that the phase transitions here are governed by subtle features of mobility edges and ``spectral inversion" which are possibly unique to structurally amorphous systems. Our results provide a concrete setting to understand the  general underpinnings of such phases -- where strong structural disorder interplays with symmetry-protected topological order. 
 \end{abstract}

\maketitle


{\it Introduction :} Symmetry protected topological (SPT) phases\cite{Wen_RMP_2017,Ludwig_PS_2015, Chiu_RMP_2016,Hasan_RMP_2010, Qi_RMP_2011, Ando_JPSJ_2013,Yang_NatMat_2012,Vergniory_arXiv_2018} of electrons, even in the weak coupling limit, in amorphous and structurally glassy systems  provide for  the interplay of structural disorder, residual global symmetries and patterns of quantum entanglement on the many-body electronic states.\cite{Agarwala_PRL_2017, Mitchell_NP_2018}
A particularly interesting and recent setting to explore such physics occur in the three dimensional $Z_2$ free fermion SPTs ({\it e.g.}, the topological band insulators (TBI)), protected by time-reversal symmetry (TRS) and particle number conservation, in simple hopping models on a structurally amorphous network.\cite{Agarwala_PRL_2017}

\begin{figure}
	\centering
	\includegraphics[width=0.85\columnwidth]{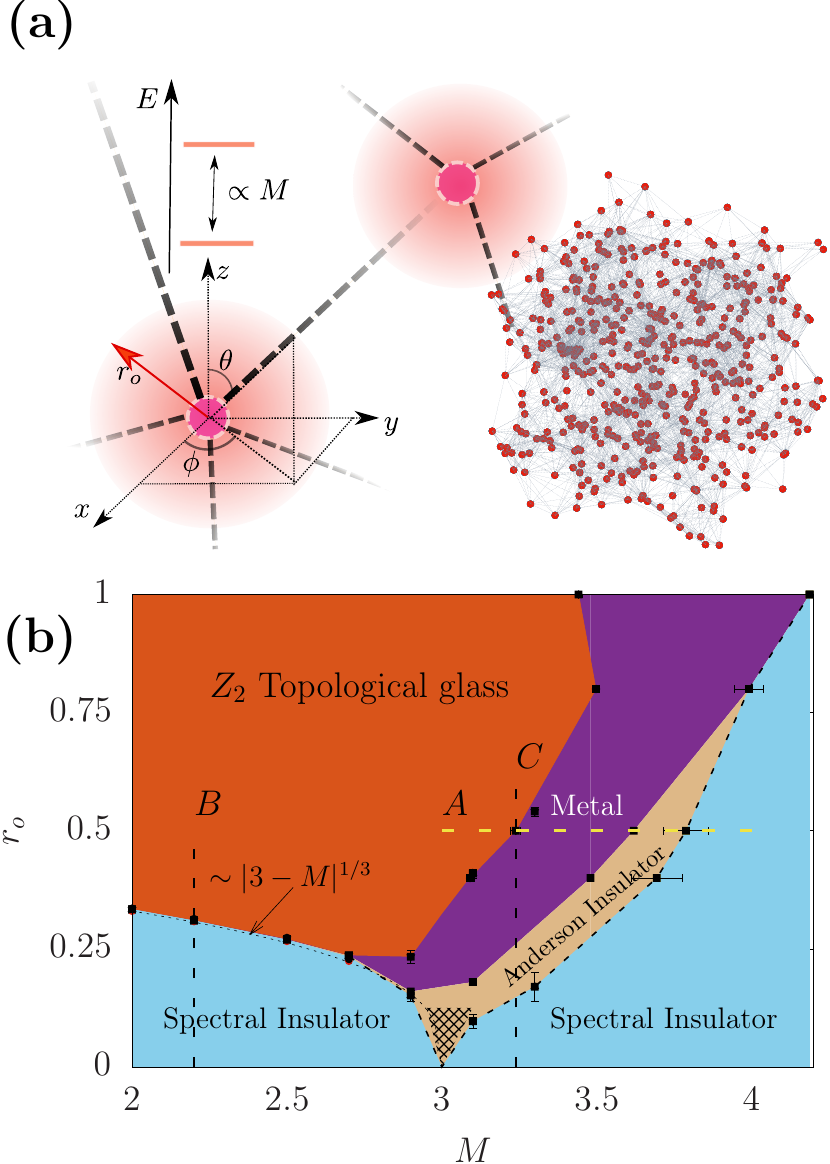}
	\caption{{\bf Phase Diagram:} (a) A model of amorphous glass is shown with a random distribution of sites in three dimensions as introduced in \cite{Agarwala_PRL_2017}. While $r_o$ characterises the ``Bohr's radius" of the atoms, $M$ is the proportional to the atomic spectral gap within the  two atomic orbitals per site (see text). (b) Phase diagram for a half-filled system in the $r_o- M$ parameter space. (A-C) are parameter cuts discussed later in the text.}
	\label{fig:PhaseDigram}
\end{figure}

The role of potential disorder in three dimensional $Z_2$ TBI and associated disorder driven quantum have transitions have so far almost exclusively investigated using a variety of methods\cite{Shindou_PRB_2009,Leung_PRB_2012,Kobayashi_PRL_2013,Song_PRB_2014,Ryu_PRB_2012,Sbierski_PRB_2014,Ohtsuki_JPSJ_2017,Akagi_JPSJ_2017} starting with an underlying crystal. Interestingly in this context Ref.~\onlinecite{Guo_PRL_2010} pointed out that even a clean metallic system can be driven by potential disorder to show topological physics, albeit in a narrow parameter regime. However, possibilities of realizing such topological phases \cite{Agarwala_PRL_2017} in three dimensional {\it electron glass}, {\it i.e.,} in a completely random network,  provide a distinct new setting to access to the underlying issues related to the characterization of such SPTs intrinsically beyond band theory. \cite{Prodan_2017, Huang_PRL_2018, Li_arXiv_2019, Bourne_RMP_2016, Loring_AOP_2015, Prodan2016bulk, Essin_PRB_2007, Prodan_JFA_2016, Jian_PRL_2009, Loring_EPL_2010, Wang_PRX_2014, Guo_PRB_2010, Varjas_arXiv_2019, Shem_arXiv_2019,Yamakage_PRB_2013, Banerjee_NS_2017, Roy_PRL_2017, Goswami_PRB_2017, Liu_PRL_2017, Loring_arXiv_2018, Chern_EPL_2019, Vittorio_JMP_2018} This, in turn, allows one to explore  questions related to novel  quantum phase transitions driven by structural disorder \cite{Lee_RMP_1985, Edwards_1972, Thouless_1974, Kramer_RPP_1993, Evers_RMP_2008} in these amorphous systems. Indeed recent extensive numerical calculations  \cite{Sahlberg_arXiv_2019} show various features-- mostly related to the nature of quantum phase transitions-- in a number of two dimensional topological phases in structurally disordered systems that are distinct from conventional crystalline systems with potential disorder.

This calls for a comprehensive characterization of electronic topological phases and phase transitions in three dimensional amorphous networks which is presently missing. In particular, we pose the question -- given a random set of atomic sites -- with no semblance of a lattice -- what generic (topological) phases can exist in three dimensions and what can be generically said about the nature of associated phase transitions out of such a phase ? Indeed the possibility of realizing such electronic phases in structurally glassy systems opens up a plethora of new questions related to their stability to interactions and slow relaxation dynamics of the underlying network. 

 In this paper, we take important steps to address both these questions in context of a three dimensional $Z_2$ topological glass (TG) using a combination of extensive numerical diagonalisation of the Hamiltonian (see below) and ideas from localisation physics. We perform extensive characterization of a three dimensional amorphous $Z_2$ free fermionic SPT through its characteristic quantized electro-magnetic response-- the so called Witten effect \cite{Witten_PLB_1979, Wilczek_PRL_1987, Rosenberg_PRB_2010}-- which serves as an effective diagnostic of the non-trivial electronic state.  While such an effect is present in three dimensional TBIs where it has been analyzed \cite{Rosenberg_PRB_2010} and corroborates with band topology based calculation of $Z_2$ invariants; in absence of the latter for e.g.~in case of TG, the detection of Witten effect and its stability as well as its annihilation becomes central to characterizing the phase. We further study the complete phase diagram to understand the possible phase transitions out of this SPT phase.  Our central result is shown in Fig. \ref{fig:PhaseDigram} where our numerical calculations show that by varying the microscopic parameters (discussed below) we can access two generic classes of transitions out of the TG-- (1) to a metallic phase which subsequently becomes an Anderson insulator with or without a spectral gap, and, (2) direct transition to an Anderson insulator with a spectral gap which we refer to as a trivial spectral insulator (SI) in the rest of this paper. Interestingly, the SI of the present work is the same state as the extensively studied structural glass with electronic spectral gaps\cite{Heine_JPC_1971,Wearie_PRB_1971,Thorpe_PRB_1971,Elliott_RMP_1974, Schwartz_PRB_1972, Davies_PRB_1984}. We provide a descriptive understanding of such transitions and via the behavior of mobility edges ideas of ``spectral" inversion.

\begin{figure}
\centering
\includegraphics[width=0.9\columnwidth]{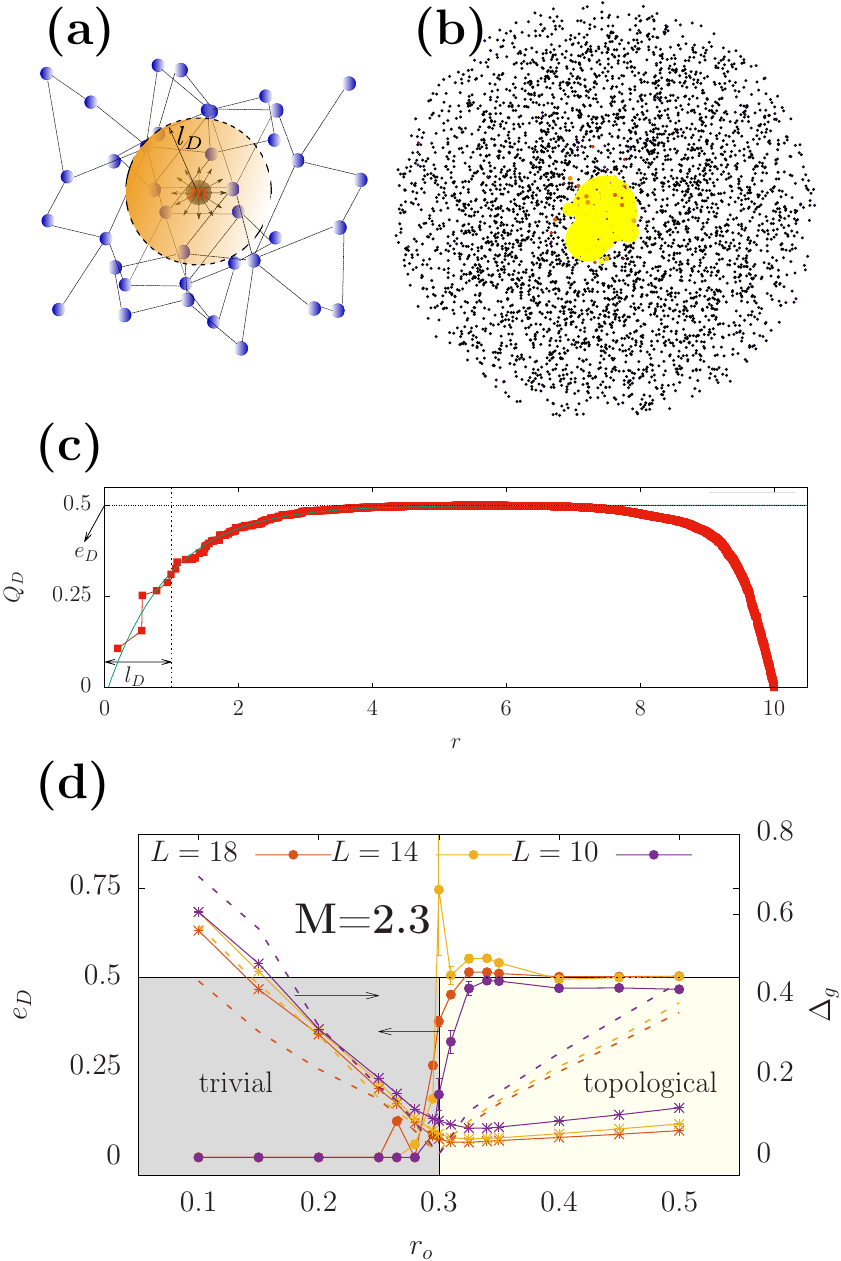}
\caption{{\bf Witten effect:} (a-b) Inclusion of a magnetic monopole (see schematic) leads to a resdistribution of the charge through out the amorphous network. (c)For $M=0.0$ and $r_0=1.0$ the value of the accumulated charge is plotted as a function of distance $r$ showing that the charge saturates to a value of $e_D \sim 1/2$. (d) Variation of the dyon charge $e_D$ as a function of $r_o$ showing a transition from a trivial spectral insulator to a topological glass phase. The corresponding spectral gap ($\Delta_g$) for both the amorphous sphere (star with solid lines) and for an amorphous cubic system with periodic boundary condition (dotted lines) is shown signaling a bulk gap closing.}
\label{fig:dyon}
\end{figure}

\paragraph*{Model for amorphous $Z_2$ insulator in three spatial dimensions :} Our starting point is the the hopping Hamiltonian introduced in Ref. \onlinecite{Agarwala_PRL_2017} :
\beq
H = \sum_{ I \alpha\sigma } \sum_{ J \beta\sigma' } t^{\sigma\sigma'}_{\alpha\beta}(\br_{IJ}) C_{I,\alpha,\sigma}^\dagger C_{J,\beta,\sigma'}
\label{eq:Ham}
\eeq
where $C_{I,\alpha,\sigma}$ is the annihilation operator for a pair (orbitals, $\alpha,\beta=s,p$) of  spin-$1/2$ ($\sigma,\sigma'=\uparrow,\downarrow$) fermionic Kramers doublets at every site (labelled by $I, J$) of a random network of ``atoms" embedded in three spatial dimensions, as is schematically shown in Fig. \ref{fig:PhaseDigram}. $\br_{IJ}$ is the vector from site $I$ to site $J$ . Centrally, for $I\neq J$, the hopping amplitude has the form $t^{\sigma \sigma'}_{\alpha\beta}(\br)=t(r)T^{\sigma \sigma'}_{\alpha\beta}(\hat {\bf r}) $ where $T^{\sigma \sigma'}_{\alpha\beta}(\hat {\bf r})$ captures the angular dependence of hopping  (see Appendix~\ref{App:HamSym}) and $t(r)=\Theta(R-r)e^{-(r-r_o)/r_o}$ determines the distance dependence. Further, given number of sites, $ N$, embedded in a spatial volume $V=L^3$ ($L$ is the linear dimension), the essential length scale of average distance between sites, $a_0= (\frac{V}{N})^{1/3}$ is set to $1$ in the rest of the paper. Finally, we consider the system at half-filling, i.e.~the number of total electrons are $2N$.

In this unit, $t(r)$ encodes two comparative length scales-- $r_o$ and $R$. The former, ``a Bohr's radius" quantifies the essential size of the atomic orbitals while the latter enforces a hard-cut off that keeps the hopping short-ranged even when $r_o$ is tuned. Note that, it is important to keep $R\ll L$ in order to treat the system as three dimensional all the way up to $r_0\sim R$. Indeed for $r_0\sim R\sim L$ the system reduces to a zero dimensional ``quantum dot" with many {\it internal} states. The structure of $t^{\sigma \sigma'}_{\alpha\beta}(0) \equiv \epsilon_{\alpha \beta}$ dictates the onsite energy of the system which is tuned by a single ``mass" parameter $M$ (See Appendix~\ref{App:HamSym}) . Thus, for a given $R$, both $r_0$ and $M$ determines the phase of the system which can then be varried to yield the phase diagram (fig. \ref{fig:PhaseDigram}). Broadly, $M$ controls the spectral gap while $r_o$ governs the strength of hopping between randomly placed atoms within a sphere of radius $R$-- hence simultaneously controlling the physics of delocalization and as well as effective disorder. Indeed, Ref.~\onlinecite{Agarwala_PRL_2017} showed the existence of delocalized surface states for a particular set of parameter in this system, suggesting a $Z_2$ topological glass phase. 

{
\paragraph*{The Witten Effect :} 
A comprehensive characterisation of the $Z_2$ insulator in three dimensions can be obtained by the Witten effect. This can be understood from the effective low energy Lagrangian of a time-reversal symmetric insulator which contains an axion contribution of $\Delta {\cal L} = \frac{\alpha}{4 \pi^2} \theta \textbf{E}.\textbf{B}$ ($\alpha$ being the fine-structure constant) in addition to the usual Maxwell action. Here, the axion angle, $\theta$ is $0 (mod(2\pi))$ (trivial SI) or $\pi(mod(2\pi))$ (TG) due to TRS.  Due to the non-zero axion angle, when a magnetic monopole ($m$) is placed inside a $Z_2$ insulator, it binds a half odd integer electric charge  ($e$) with it to form a {\it dyon} \cite{Saha_PR_1949, Wilczek_PRL_1987, Witten_PLB_1979}.  Indeed properties of such {\it dyons} can lead to identification of new SPT phases in presence of interactions. \cite{Metlitski_PRB_2013, Wang_Science_2014, Metlitski_PRB_2016, Maciejko_NP_2015, Maciejko_PRL_2010, Swingle_PRB_2011, Bhattacharjee_PRB_2012}

 To investigate this effect in a topological glass, we introduce an unit magnetic monopole (see Appendix~\ref{App:Witten}) which couple to the electrons through minimal coupling $t_{\alpha\beta}({\bf r}_{IJ})\rightarrow e^{iA_{IJ}}t_{\alpha\beta}({\bf r}_{IJ})$ where $A_{IJ}$ is the total Peierls phase between sites $I$ and $J$.  This leads to a shift in the electronic charge density at every site (see Fig~\ref{fig:dyon}). The cumulative electronic charge density inside a sphere of radius $r$ measured from the location of the monopole, $Q_D(r)$,  is also shown in Fig.~\ref{fig:dyon} for a half filled system. Noticeably, $Q_D(r)$ saturates to a value of $1/2$ within numerical accuracy illustrating realization of Witten effect in an amorphous glass. That this half of the charge had distributed from the boundary is evident by the fall of $Q_D$ to $zero$ as $r$ approaches $L$ showing that the compensating half of the charge lives in the boundary. Fitting the region of $r<L$ to a functional form of $\sim e_D(1-e^{-(r-a)/\xi})$ estimates the dyon charge $e_D$ and corresponding ``size" of the dyon $\equiv l_D=\xi+a$ (see \Fig{fig:dyon}). However, the value of dyon length and the way $e_D$ transits from $0$ to $0.5$ (modulo $1$) is configuration sensitive (see Appendix.~\ref{App:Witten}).
 
 \paragraph*{The phase diagram :} 
 
The Witten effect not only characterizes the TG, but as shown in Fig.~\ref{fig:dyon}, the electric charge of the dyon disappears across transition to a trivial spectral insulator. Not surprisingly, the disappearance of the Witten effect coincides with the closing of the bulk spectral gap for a system with periodic boundary conditions as is evident from Fig.~\ref{fig:dyon}. Such a bulk gap closing is essential for the the axion angle to change and thereby signaling the phase transition. This immediately raises the general question of the characterization of nature of the phases and associated transitions out of the TG in the whole $(r_0, M)$ plane.  We perform exact diagonalizations over $\sim 200$ configuration realizations for system sizes $L=8-14$ to calculate, in addition the to the Witten effect, the spectral gap, the inverse participation ratio (IPR) and the orbital nature of the states close to the Fermi level to obtain the phase diagram plotted in Fig. \ref{fig:PhaseDigram}. To implement periodic boundary condition in any direction we allow fermionic hopping in a ``repeated" configuration with distances modulo $L$ in all directions and identify equivalent sites.

 To this end, let us concentrate on a typical cut shown as line  A in Fig. \ref{fig:PhaseDigram} at a constant $r_0=0.5$ as a function of $M$ and consider states within a small window of states ($\delta(E)=10\times E_{bw}/(4 N)$ also shown in Fig.~\ref{fig3}(a)) around the Fermi level, where $E_{bw}$ is the bandwidth of the spectrum. In Fig.~\ref{fig3}(a), we plot the configuration averaged  spectral gap as a function of $M$. The finite spectral gap region for $M\lesssim 3.2$ is the $Z_2$ topological glass phase with axion angle $\theta=\pi$ leading to half quantized Witten effect discussed above while for the regime $3.2\lesssim M\lesssim 3.8$ where the spectral gap is zero (within errorbars) represents a gapless phase. Beyond $M\gtrsim 3.8$, the spectral gap reopens and this is a trivial spectral insulator with axion angle $\theta=0$ (not shown). 
 
    \begin{figure}
\includegraphics[width=0.85\columnwidth]{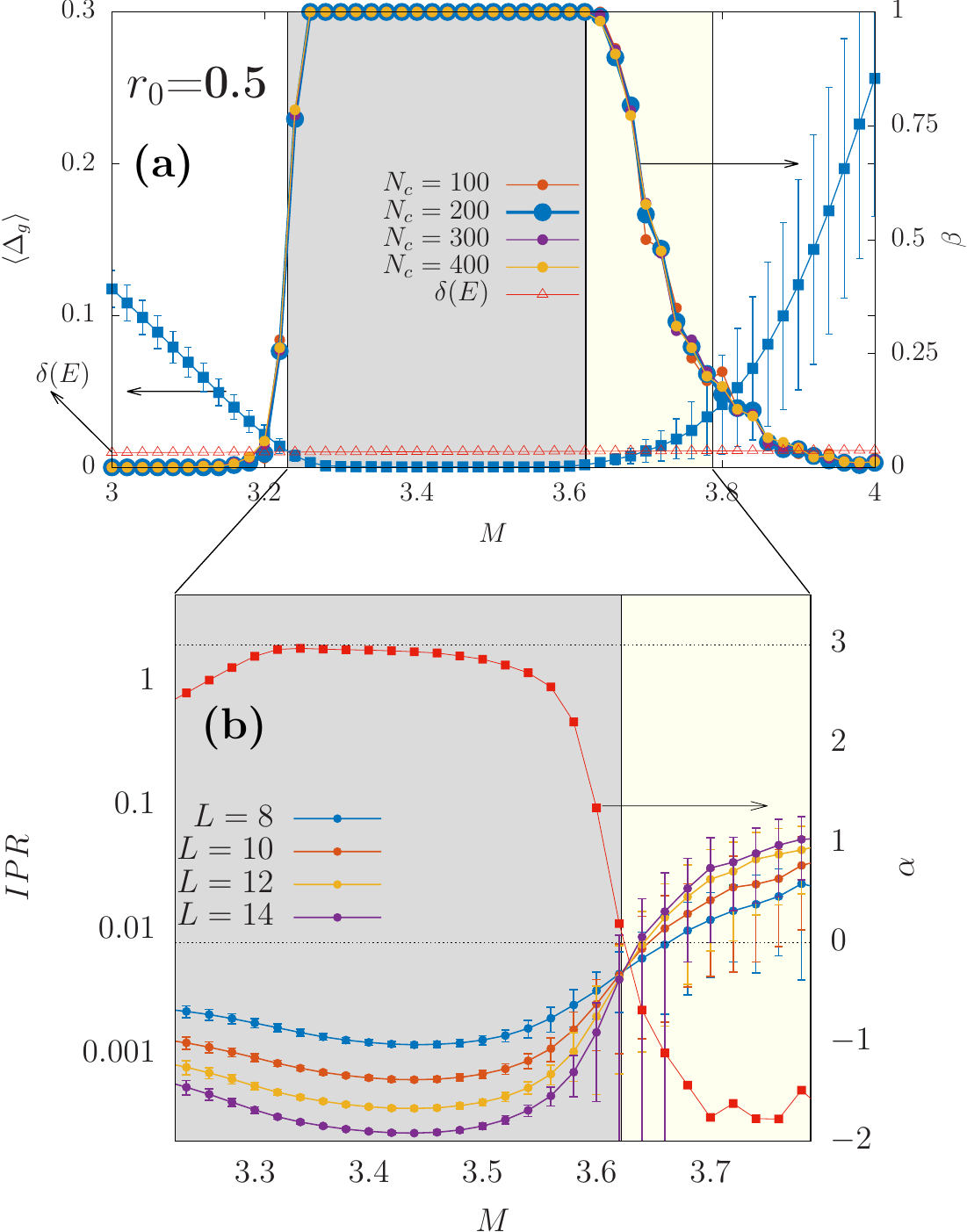}
\caption{{\bf Intermediate phases:} (a) Configuration averaged spectral gap $\Delta_g$ as a function of $M$. Also shown is the fraction of configurations which have energy states within a small energy window $\delta(E)$, defined as $\beta$ for different number of configurations. (b) Apart from the two gapped phases, one obtains a metallic phase where IPR of the states scale as $1/L^\alpha$ with $\alpha \sim 3$; variation of $\alpha$ is shown. The rare-region dominated Anderson localization shows anomalous scaling with $\alpha<0$ ($N_c=200$) . }
\label{fig3}
\end{figure}

 At this point it is important to note that the for a finite, $N_c=100-400$, number of independent configurations, we can calculate the subset of configurations, $N_I$, that lead to at least one state within the above energy window of $\delta(E)$. The fraction of such states, $\beta = N_I/N_c$, is also plotted in Fig. \ref{fig3}(a) and we find that while for the transition from the $Z_2$ topological glass to gapless phase, the fraction quickly goes to one from zero, the change is much more gradual for the transition from gapless phase to the trivial spectral insulator on the other side. In parallel, our data suggest stark asymmetry of the error bars on the value of the spectral gap on either side of the gapless window. While the error-bars are much smaller on the $Z_2$ topological glass side or on the gapless region, it is much bigger on the other end suggesting that the spectral gap has a wider distribution near the transition between the gapless state and the trivial spectral insulator (see Appendix.~\ref{App:IPR}).  This observation is further reinforced by the analysis of the configuration averaged IPR in the gapless regime ($3.2\lesssim M\lesssim 3.8$) as shown in Fig.~\ref{fig3}(b).  We find that  configuration averaged typical states in the parameter range $3.2\lesssim M\lesssim 3.6$ has an IPR that scales with system as  $\sim 1/L^{\alpha}$ where $\alpha \sim 3$ (see Fig.~\ref{fig3}) -- characteristic of a {\it metallic} phase. However, in the narrow window $3.6\lesssim M\lesssim 3.8$, the IPR behaviour is consistent with a negative $\alpha$ when fitted to the above form with $M=3.6$ being the point where $\alpha=0$ and changes sign and becomes negative! Indeed, it is interesting to note the prominent crossing of IPRs of the different system sizes at $M=3.6$. Again there are large error bars in the regime $M\gtrsim 3.6$ where $\alpha$ is close to zero. This is typical of formation of localized states near a metal-Anderson Insulator transition suggesting the onset of Anderson localization \cite{Lee_RMP_1985, Kramer_RPP_1993}. 
 
 In the lights of possible localization physics at play, the negative $\alpha$ finds a natural explanation. That with increasing system size, probability of finding {\it more} dominantly localized states increases -- is a feature, attributed to rare regions that are known to occur in the Liftshitz tails of a disordered band. Given the two band character of our energy spectrum -- the states near the Fermi energy are indeed these rare regions belonging to both the upper and lower band -- thereby providing for the surprising feature in the IPR as mentioned above.  With further increase in $M$ even the Lifshitz tails move beyond the band spectrum -- leading to a trivial spectral insulator as seen in \Fig{fig3}. 
  
  In order to further understand the general phases and their transitions, we now investigate the orbital behavior and  the energy spectrum of a single amorphous configuration as a function of $r_o$ for two cuts (B and C) as shown in (see \Fig{fig:PhaseDigram}(b)). Both these vertical cuts have a simple trivial atomic insulator limit when $r_0=0$ where the network degenerates to disconnected atoms with the on-site {\it atomic orbitals} leading to two isolated spin degenerate states on each atom separated by an onsite energy $|3-M|$ in the present parametrization. At half filling, this gives a trivial atomic insulator (unless $M=3$ where it is a highly degenerate state susceptible to perturbations leading to the breakdown of our numerical accuracy as mentioned in Fig. \ref{fig:PhaseDigram}.).  For a fixed finite $R$, tuning $r_o>0$ (at a constant $M$) modulate the strength of the connectivity of the electronic orbitals on different atoms  of the disordered network (see Appendix~\ref{App:Disorder}). While the trivial atomic insulator is stable to small $r_0>0$, the energy eigenstates are no longer strictly onsite leading to the formation of two bands of fully localized states around energies, $\pm |3-M|$. On further increasing $r_0$ the states in the middle of each of these bands become delocalized giving rise  to two pairs of mobility edges which approaches the fermi level with increasing $r_0$ and eventually leading to the closing of the spectral gap at a critical $r_0=r_{0B(C)}$ fo the cut B(C). This is shown in Fig. \ref{fig4}(b). The extent of the delocalization of the states near the Fermi level is characterized by their IPR which is shown in Appendix~\ref{App:IPR}.

\begin{figure}
	\includegraphics[width=1.0\columnwidth]{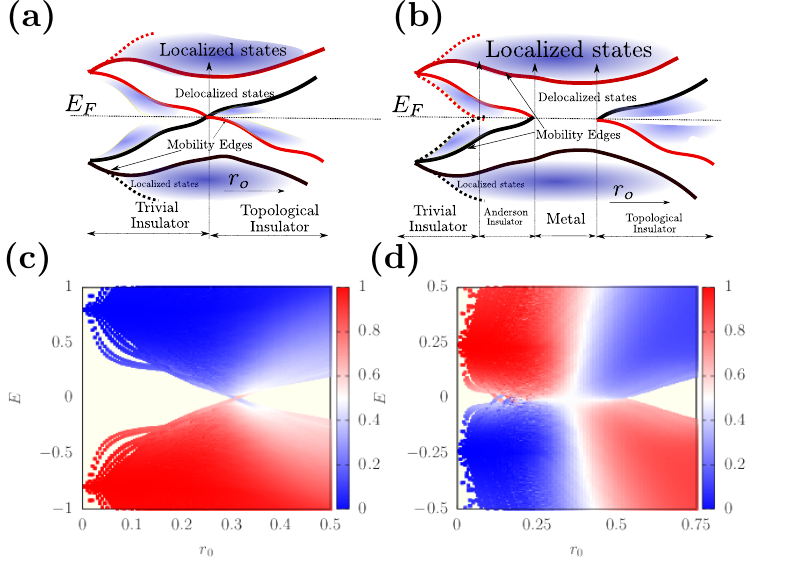}
	\caption{{\bf Phase transitions:} (a) Schematic showing that the transition across the cut B ($M=2.2$) and (b) across cut C ($M=3.24$) of the phase diagram as shown in \Fig{fig:PhaseDigram}. The schematic shows the various intermediate phases which are encountered. (c) and (d) is the behavior of the energy spectrum as a function of $r_o$ for a particular periodic configuration ($R=2,L=14$). The color bar shows the $s$ orbital character of each of the state. While in cut B it is the highest occupied orbital which contributes to shifting of the spectral weight -- in cut C many low energy states contribute to this feature due to an intervening metallic phase.}
	\label{fig4}
\end{figure}

For cut B, the spectral gap immediately (within our numerical resolution) reopens on cross $r_{0B}$. However the orbital nature of the occupied orbitals change across the bulk gap closing transition. In this sense, we have an inverted spectral gap for $r_0>r_{0B}$ (compared to the atomic insulator) which is accompanied by winding of the axion angle from $0$ to $\pi$ leading to the half-quantized Witten effect observed before. This is nothing but the $Z_2$ topological glass. For the cut C, on the other hand, the situation is more interesting and involved. While, not surprisingly, again the trivial atomic insulator is stable to very small $r_0>0$, the spectral gap closes as $r_0$ is increased. However, the IPR calculations show that the states are not extended (see Appendix.~\ref{App:Disorder}). This therefore is a gapless localized phase similar to an Anderson Insulator. On increasing $r_0$ further, the mobility edges which had developed inside the two 	``bands" meet at the fermi level leading to a metallic state with delocalised bulk states at the Fermi level as confirmed by our IPR results. This metallic phase thus has finite conductance.  It is interesting to note that the orbital resolved spectral weights of the change their character inside the metal (fig.~\ref{fig4}) such that when the spectral gap opens up on further increase of $r_0$ we get the inverted gap and we are back in the $Z_2$ topological glass.

\paragraph*{Associated phase transitions :} 
We now briefly discuss the nature of phase transitions in this system for cut B and C -- where $r_o$ is smoothly increased. Note that this does not change the connectivity, or the coordination number of the sites in a particular configuration. However, increasing $r_0$ leads to the increase in the strength of hopping compared to the mass scale $M$ and this destabilises the trivial spectral insulator when the hopping energy-scale becomes comparable to the mass scale (see the schematic \Fig{fig4}(a)). A naive estimate of the scaling relation is obtained by comparing the hopping amplitude integrated over a sphere of radius $R$ with the mass scale $|3-M|$. This therefore opens up a fan around $M=3$ such that below the scale $r_o\sim |3-M|^{1/3}$ (for $R\gg r_0$, see dashed line in \Fig{fig:PhaseDigram}) such that below this scale, trivial spectral insulator survives as a stable phase and gives way to other phases above it.

Moving out of the $Z_2$ insulator, our numerical calculations reveal that the bulk gap closing leads to either a metal or a trivial spectral insulator. For $M<3$ we find a direct transition between the a trivial and topological phases {\it without} an intermediate metallic phase (upto our numerical accuracy). As shown in Fig. \ref{fig4}, this occurs due to immediate opening up of the spectral gap along with the concomitant inversion of the orbital nature of the highest occupied state. This closure of the gap and the associated untwisting of the wave-function is then similar to the Plateau transition in the Integer Quantum Hall effect \cite{Huckestein_RMP_1995, Prange_1987}.  In the majority of the region $M>3$, however, the trivial insulator mostly gives way to a gapless phase which is either a metal with delocalised electronic states near the fermi-level or a novel Anderson Insulator with properties dominated by rare region physics.  We find that this transition is rather generically expected for an amorphous network. To check this we investigate the physics of a different ``control" Hamiltonian where $T_{\alpha\beta} $ is changed to a trivial form keeping the onsite term unchanged and keeping the time-reversal symmetry intact. This model in a crystalline limit hosts no topological phases, and in an amorphous setting has a transition at $M=3, r_o=0$ between two atomic limits. This transition fan outs to a metallic phase at finite $r_o$ near $M=3$ (see Appendix~\ref{App:Controlcase}).} However in contrast to the this control Hamiltonian our gapless metallic phase hosts a characteristic shift in the orbital character of the states near the Fermi energy -- in order to open up into a topologically nontrivial phase (see \Fig{fig4}(b)).

\paragraph*{Summary and outlook :}

We have investigated the nature of phases and phase transitions for a model of a hopping Hamiltonian (introduced in Ref. \onlinecite{Agarwala_PRL_2017}) which can host a topologically glassy phase. We have used the effective hopping range $r_o$ and the onsite orbital energies ($M$) as effective parameters to explore the phase space. We find a rich phase diagram with a wide regime of the system being in the topological regime -- and interesting intermediate phases and phase transitions. We characterise the system using exact diagonalization techniques and by investigating the properties of the wavefunctions. In particular we characterize the topological (and trivial) phase through the occurence (or absence) of Witten effect. We build a comprehensive understanding of the phase diagram by alluding to ideas in the disorder literature and by examining the orbital character of the occupied orbitals in this system. We further point out that few transitions in this system could be of an distinct character than those Anderson transitions which could otherwise be found in amorphous networks.  It would be an interesting future direction to  completely characterize such transitions and their dependence on the symmetries, depending if they belong to a particular Altland-Zirnbauer classes of the tenfold way \cite{Chiu_RMP_2016, Ludwig_PS_2015,Ryu_NJP_2010, Agarwala_AOP_2017}.

{\it Acknowledgements:}
We acknowledge fruitful discussions with Vijay B. Shenoy, Sumilan Banerjee. AA acknowledges the many enlightening discussions with Vijay~B. Shenoy and the particular collaboration\cite{Agarwala_PRL_2017} where many of these ideas had germinated. PM acknowledges funding from CSIR, India, through Shyama Prasad Mukherjee Fellowship.
 AA and SB acknowledge funding from Max Planck Partner Grant at ICTS.  SB acknowledges SERB-DST (India) for funding through project grant No. ECR/2017/000504. Numerical calculations were performed on clusters {\it boson} and {\it tetris} at ICTS.

\appendix

\begin{figure}
	\includegraphics[width=1.0\columnwidth]{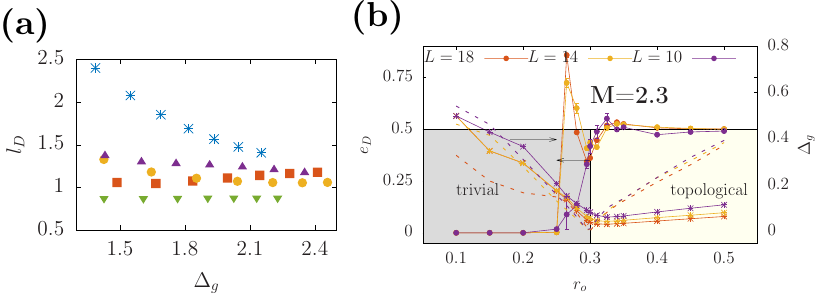}
	\caption{{\bf Dyon length:} Variation of dyon length ($l_D$) for a sphere of amorphous TI and the corresponding spectral gap $\Delta_g$ when it is embedded as the periodic cubic amorphous TI for different choice of configurations (denoted by different point types) for various choices of $\{M,r_o\}$, broadly showing that dyon length falls with increasing band gap ($L=16$). Note that $l_D \ll L$.}
	\label{supfig1}
\end{figure}

\section{Hamiltonian and Symmetries}
\label{App:HamSym}
Following \cite{Agarwala_PRL_2017} the form of the Hamiltonian is given by Diag$\{-3+M,-3+M,3-M,3-M\}$
and
{\small
	\begin{equation}
	T^{\sigma \sigma'}_{\alpha\beta}(\hat r) = \left( \begin{array}{cccc} 
	1 & 0 & -i\cos{\theta} & -i e^{-i\phi}\sin{\theta}
	\\0 & 1 & -i e^{i\phi}\sin{\theta} & i\cos{\theta} 
	\\ -i\cos{\theta} & -i e^{-i\phi}\sin{\theta} & -1 & 0
	\\-i e^{i\phi}\sin{\theta} & i\cos{\theta} & 0& -1
	\end{array} \right) 
	\label{hammat}
	\end{equation}}
where $\theta$ and $\phi$ are polar and azimuthal angle respectively for $\br_{IJ}$, which is the relative vector from site $I$ to site $J$(see \Fig{fig:PhaseDigram}(a)). More generally the structure of $T^{\sigma \sigma'}_{\alpha\beta}$ is constrained due to symmetries. For instance time reversal which is an antiunitary symemtry that acts on the fermionic operators as ${\cal T} c_{I \sigma}  {\cal T}^{-1} = \sigma c_{I\sigma}$ constraints $T_{\alpha \beta} ^{\sigma \sigma^\prime} (\theta , \phi) = [T_{\alpha \beta} ^{\bar{\sigma} \bar{\sigma}^\prime} (\theta , \phi)]^* \bar{\sigma} \bar{\sigma}^\prime $. Moreover hermiticity demands $T_{\alpha \beta} ^{\sigma \sigma^\prime} (\theta , \phi) = [T_{\alpha \beta} ^{\sigma \sigma^\prime} (\pi - \theta , \phi - \pi)]^\dagger $.

\begin{figure}
	\includegraphics[width=0.95\columnwidth]{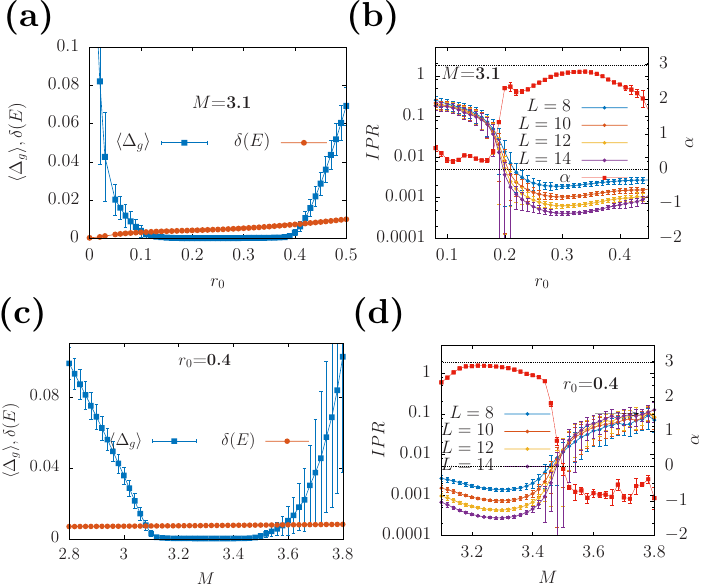}
	\caption{{\bf Metal-Insulator transitions:} Configuration averaged spectral gap $\Delta_g$ and $\delta(E)$ (see text) as a function of (a) $r_o$ at $M=3.1$ and (c) $M$ for $r_o=0.4$. IPR and their scaling exponent with system size ($\alpha$) for the states near the Fermi energy are correspondingly given in $(c)$ and $(d)$ ($N_c=200$).}
	\label{supfig2}
\end{figure}

\begin{figure}
	\includegraphics[width=0.95\columnwidth]{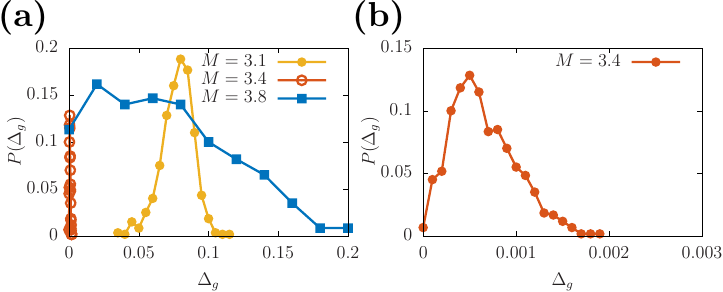}
	\caption{{\bf Distribution of spectral gap:} Regions in the parameter space have different distribution of spectral gaps. (a) For  $r_o=0.5, M=3.1,3.4,3.8$ are shown. In the metallic phase i.e.~at $M=3.4$ the distribution is narrow which is zoomed in (b) ($R=2,L=10, N_c=600$).}
	\label{supfig3}
\end{figure}

\section{Details about the Witten effect}
\label{App:Witten}

In order to observe the Witten effect the following protocol is followed. A magnetic monopole, corresponding to a vector potential $\bA = \frac{-g}{r\sin{\theta}} (1+\cos{\theta}) \hat{\phi}$, is placed at the center of an amorphous glass sphere of diameter $L$ (as shown in \Fig{fig:dyon}).  $g$ is the magnitude of the monopole strength which we fix to $1$.  This leads to additional Peirels phase terms on the different hoppings (see \eqn{eq:Ham}). Local electronic charge density is calculated for a half-filled system both with($m=1$) and without($m=0$) the magnetic monopole.  The local difference in charge density at site $I$ is given by $\Delta \rho(\br_I) = \rho(\br_I)_{m=1}-\rho(\br_I)_{m=0}$ where $\rho(\br_I)_{m=0(1)} = \sum_{n \in occ} \sum_{\alpha, \sigma} |\psi^{m=0(1)}_{n}(I, \alpha~\sigma)|^2$. $n$ labels all the eigenstates which are occupied, $\alpha$ is the spin/orbital index and $\br_I$ is the real space position vector of site $I$ where the redistributed charge density is evaluated. A finite sized system leads to inadvertent hybridization of boundary modes, in which case it is technically prudent to keep one additional state (above the occupied states) in calculation of density for both with and without the monopole. In order to calculate the effective change in charge density around the point where the magnetic monopole is introduced, one can evaluate the effective electric charge around the monopole using $Q_D(r) = \int_{|\br'|=0}^{r} d^3\br' \Delta \rho(\br') = \sum_{I, \br_I \leq r} \Delta \rho(\br_I)$ where $r$ is the distance from the origin. A variation of $Q_D(r)$ as a function of $r$ is shown in \Fig{fig:dyon}. We remark here that while existence of Witten effect, by itself occurs rather sharply, the dyon length (see \Fig{fig:dyon}(b)) is configuration dependent. We observe that generally dyon length is smaller if the corresponding periodic system has a larger spectral gap $(\Delta_g)$ (see \Fig{supfig1}). This is expected behaviour since dyon length is expected to diverge as we go close to a quantum phase transition where the spectral gap goes to zero.

\begin{figure}
	\includegraphics[width=0.95\columnwidth]{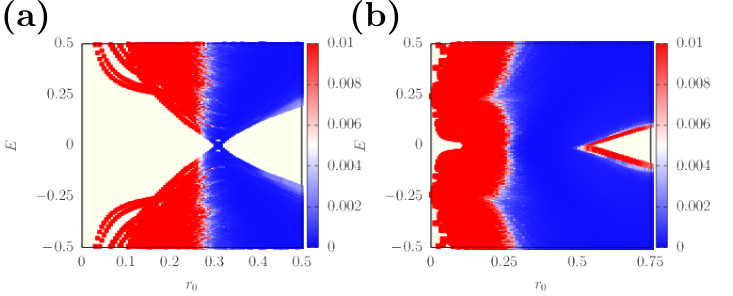}
	\caption{{\bf Variation of IPR:} 
		(a) and (b) is the behavior of the energy spectrum as a function of $r_o$ for a particular periodic configuration at $M=2.2$ and $M=3.24$. The color bar shows the IPR of each of the  the state ($R=2,L=14$).}
	\label{supfig3b}
\end{figure}

\begin{figure}
	\includegraphics[width=0.9\columnwidth]{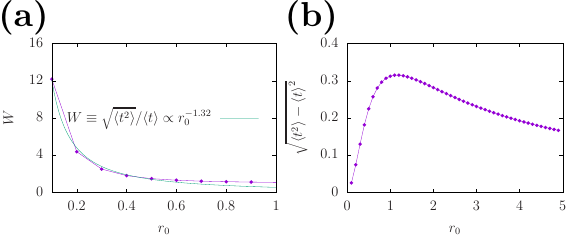}
	\caption{{\bf Disorder estimate:} Behavior of mean, standard deviation of $t(r)$ of an amorphous network ($R=2$) and the effective disorder scale ($\equiv W$) for an amorphous network in $L=16^3$ system as a function of $r_o$. }
	\label{supfig4}
\end{figure}

\section{IPR and Spectral gap}
\label{App:IPR}

At any value of $(M,r_o)$ one can calculate the configuration averaged spectral gap and the IPR of the states near the Fermi energy, if any. The behavior of such states were shown in \Fig{fig3}. We provide further results for two different parameter regimes in \Fig{supfig2}. We further point out that the distribution of the spectral gap depends on which phase we are in. For instance in near to rare-region dominated SI the distribution of spectral gaps is extremely wide (see \Fig{supfig3}). We also show the IPR of the energy spectrum and its behavior with $r_o$ corresponding to cut B and cut C (as discussed in \Fig{fig:PhaseDigram} and \Fig{fig4}) in \Fig{supfig3b}.

\begin{figure}
	\includegraphics[width=0.9\columnwidth]{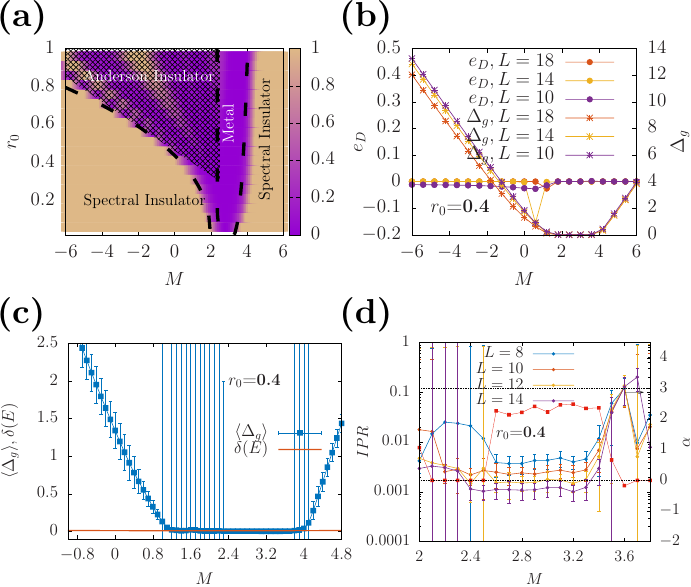}
	\caption{{\bf Control case:} (a) Configuration averaged spectral gap in the $M-r_o$ plane for $20$ configurations($L=10$). Dashed lines are guide to eye. The behavior of the phase is shown in various regions.  Variation of spectral gap and IPR of the states near Fermi energy is shown in $(c)$ and $(d)$ as a function of $M$ for $r_0=0.4$ $(R=2)$. The large fluctuations in $\Delta_g$ and IPR are characteristic of the Anderson localized regime. Dyon charge ($e_D$) and the corresponding spectral gap for a particular amorphous sphere is shown in (b) showing absence of any topological regime.}
	\label{supfig5}
\end{figure} 

\section{Disorder Estimate}
\label{App:Disorder}

An amorphous network, as discussed here, incorporates two ``kinds" of disorder scale. One -- in terms of random connectivity and therefore the fluctuation of coordination number of the sites. The second, as the distribution of  hopping strengths themselves. While $R$ decides the former, $r_o$ determines the latter (see discussion near \eqn{eq:Ham}). Therefore keeping $R$ fixed one can tune $r_o$ to change the effective disorder scale in the system $W \equiv \sqrt{\langle t^2 \rangle}/\langle t \rangle \propto r^{-3/2}_o $. For a given configuration we numerically calculate this value whose behavior is shown in \Fig{supfig4}. A sharp upward rise in effective $W$ is noticed in the range of $r_o\sim 0.25$ which is the scale where complete localization of all the states can be expected.

\section{Control case}
\label{App:Controlcase}

Note that \eqn{hammat} when implemented in a cubic crystalline system leads to $H = (\bOne \otimes \tau_z) [-3 + M + \cos{k_x} + \cos{k_y} +\cos{k_z}] + 
(\sigma_x \otimes \tau_x)[\sin{k_x}]+ 
(\sigma_y \otimes \tau_x)[\sin{k_y}]+ 
(\sigma_z \otimes \tau_x)[\sin{k_z}]
$ which realises a three-dimensional Dirac cone at $M=0$. We consider a deformed $H = (\bOne \otimes \tau_z) [-3 + M + \cos{k_x} + \cos{k_y} +\cos{k_z}] + 
(\sigma_x \otimes \tau_x)[\sin{k_x}+\sin{k_y} +\sin{k_z}]$. On a random lattice we generalise this to $T^{\sigma \sigma'}_{\alpha\beta}(\hat r)=-i[(\cos(\phi)+\sin(\phi))\sin(\theta)+\cos(\theta)] (\sigma_x \otimes \tau_x)$. This respects time-reversal symmetry and hermiticity. We analyze the phases of this model in the $r_o-M$ plane. We find an intermediate a metallic phase near $M\sim 3$ between two spectral insulator phases. This can be seen in \Fig{supfig5} which shows the IPR, $\Delta_g$ and Witten effect in this system for variety of parameters. Importantly this has no topological regime since the system shows no Witten effect.

\bibliography{Wittenglass}

\begin{thebibliography}{68}%
\makeatletter
\providecommand \@ifxundefined [1]{%
 \@ifx{#1\undefined}
}%
\providecommand \@ifnum [1]{%
 \ifnum #1\expandafter \@firstoftwo
 \else \expandafter \@secondoftwo
 \fi
}%
\providecommand \@ifx [1]{%
 \ifx #1\expandafter \@firstoftwo
 \else \expandafter \@secondoftwo
 \fi
}%
\providecommand \natexlab [1]{#1}%
\providecommand \enquote  [1]{``#1''}%
\providecommand \bibnamefont  [1]{#1}%
\providecommand \bibfnamefont [1]{#1}%
\providecommand \citenamefont [1]{#1}%
\providecommand \href@noop [0]{\@secondoftwo}%
\providecommand \href [0]{\begingroup \@sanitize@url \@href}%
\providecommand \@href[1]{\@@startlink{#1}\@@href}%
\providecommand \@@href[1]{\endgroup#1\@@endlink}%
\providecommand \@sanitize@url [0]{\catcode `\\12\catcode `\$12\catcode
  `\&12\catcode `\#12\catcode `\^12\catcode `\_12\catcode `\%12\relax}%
\providecommand \@@startlink[1]{}%
\providecommand \@@endlink[0]{}%
\providecommand \url  [0]{\begingroup\@sanitize@url \@url }%
\providecommand \@url [1]{\endgroup\@href {#1}{\urlprefix }}%
\providecommand \urlprefix  [0]{URL }%
\providecommand \Eprint [0]{\href }%
\providecommand \doibase [0]{http://dx.doi.org/}%
\providecommand \selectlanguage [0]{\@gobble}%
\providecommand \bibinfo  [0]{\@secondoftwo}%
\providecommand \bibfield  [0]{\@secondoftwo}%
\providecommand \translation [1]{[#1]}%
\providecommand \BibitemOpen [0]{}%
\providecommand \bibitemStop [0]{}%
\providecommand \bibitemNoStop [0]{.\EOS\space}%
\providecommand \EOS [0]{\spacefactor3000\relax}%
\providecommand \BibitemShut  [1]{\csname bibitem#1\endcsname}%
\let\auto@bib@innerbib\@empty
\bibitem [{\citenamefont {Wen}(2017)}]{Wen_RMP_2017}%
  \BibitemOpen
  \bibfield  {author} {\bibinfo {author} {\bibfnamefont {X.-G.}\ \bibnamefont
  {Wen}},\ }\href {\doibase 10.1103/RevModPhys.89.041004} {\bibfield  {journal}
  {\bibinfo  {journal} {Rev. Mod. Phys.}\ }\textbf {\bibinfo {volume} {89}},\
  \bibinfo {pages} {041004} (\bibinfo {year} {2017})}\BibitemShut {NoStop}%
\bibitem [{\citenamefont {Ludwig}(2016)}]{Ludwig_PS_2015}%
  \BibitemOpen
  \bibfield  {author} {\bibinfo {author} {\bibfnamefont {A.~W.~W.}\
  \bibnamefont {Ludwig}},\ }\href
  {http://stacks.iop.org/1402-4896/2016/i=T168/a=014001} {\bibfield  {journal}
  {\bibinfo  {journal} {Physica Scripta}\ }\textbf {\bibinfo {volume} {2016}},\
  \bibinfo {pages} {014001} (\bibinfo {year} {2016})}\BibitemShut {NoStop}%
\bibitem [{\citenamefont {Chiu}\ \emph {et~al.}(2016)\citenamefont {Chiu},
  \citenamefont {Teo}, \citenamefont {Schnyder},\ and\ \citenamefont
  {Ryu}}]{Chiu_RMP_2016}%
  \BibitemOpen
  \bibfield  {author} {\bibinfo {author} {\bibfnamefont {C.-K.}\ \bibnamefont
  {Chiu}}, \bibinfo {author} {\bibfnamefont {J.~C.~Y.}\ \bibnamefont {Teo}},
  \bibinfo {author} {\bibfnamefont {A.~P.}\ \bibnamefont {Schnyder}}, \ and\
  \bibinfo {author} {\bibfnamefont {S.}~\bibnamefont {Ryu}},\ }\href {\doibase
  10.1103/RevModPhys.88.035005} {\bibfield  {journal} {\bibinfo  {journal}
  {Rev. Mod. Phys.}\ }\textbf {\bibinfo {volume} {88}},\ \bibinfo {pages}
  {035005} (\bibinfo {year} {2016})}\BibitemShut {NoStop}%
\bibitem [{\citenamefont {Hasan}\ and\ \citenamefont
  {Kane}(2010)}]{Hasan_RMP_2010}%
  \BibitemOpen
  \bibfield  {author} {\bibinfo {author} {\bibfnamefont {M.~Z.}\ \bibnamefont
  {Hasan}}\ and\ \bibinfo {author} {\bibfnamefont {C.~L.}\ \bibnamefont
  {Kane}},\ }\href {\doibase 10.1103/RevModPhys.82.3045} {\bibfield  {journal}
  {\bibinfo  {journal} {Rev. Mod. Phys.}\ }\textbf {\bibinfo {volume} {82}},\
  \bibinfo {pages} {3045} (\bibinfo {year} {2010})}\BibitemShut {NoStop}%
\bibitem [{\citenamefont {Qi}\ and\ \citenamefont {Zhang}(2011)}]{Qi_RMP_2011}%
  \BibitemOpen
  \bibfield  {author} {\bibinfo {author} {\bibfnamefont {X.-L.}\ \bibnamefont
  {Qi}}\ and\ \bibinfo {author} {\bibfnamefont {S.-C.}\ \bibnamefont {Zhang}},\
  }\href {\doibase 10.1103/RevModPhys.83.1057} {\bibfield  {journal} {\bibinfo
  {journal} {Rev. Mod. Phys.}\ }\textbf {\bibinfo {volume} {83}},\ \bibinfo
  {pages} {1057} (\bibinfo {year} {2011})}\BibitemShut {NoStop}%
\bibitem [{\citenamefont {Ando}(2013)}]{Ando_JPSJ_2013}%
  \BibitemOpen
  \bibfield  {author} {\bibinfo {author} {\bibfnamefont {Y.}~\bibnamefont
  {Ando}},\ }\href@noop {} {\bibfield  {journal} {\bibinfo  {journal} {Journal
  of the Physical Society of Japan}\ }\textbf {\bibinfo {volume} {82}},\
  \bibinfo {pages} {102001} (\bibinfo {year} {2013})}\BibitemShut {NoStop}%
\bibitem [{\citenamefont {Yang}\ \emph {et~al.}(2012)\citenamefont {Yang},
  \citenamefont {Setyawan}, \citenamefont {Wang}, \citenamefont {{Buongiorno
  Nardelli}},\ and\ \citenamefont {Curtarolo}}]{Yang_NatMat_2012}%
  \BibitemOpen
  \bibfield  {author} {\bibinfo {author} {\bibfnamefont {K.}~\bibnamefont
  {Yang}}, \bibinfo {author} {\bibfnamefont {W.}~\bibnamefont {Setyawan}},
  \bibinfo {author} {\bibfnamefont {S.}~\bibnamefont {Wang}}, \bibinfo {author}
  {\bibfnamefont {M.}~\bibnamefont {{Buongiorno Nardelli}}}, \ and\ \bibinfo
  {author} {\bibfnamefont {S.}~\bibnamefont {Curtarolo}},\ }\href {\doibase
  10.1038/nmat3332} {\bibfield  {journal} {\bibinfo  {journal} {Nature
  Materials}\ }\textbf {\bibinfo {volume} {11}},\ \bibinfo {pages} {614}
  (\bibinfo {year} {2012})}\BibitemShut {NoStop}%
\bibitem [{\citenamefont {Vergniory}\ \emph {et~al.}(2018)\citenamefont
  {Vergniory}, \citenamefont {Elcoro}, \citenamefont {Felser}, \citenamefont
  {Bernevig},\ and\ \citenamefont {Wang}}]{Vergniory_arXiv_2018}%
  \BibitemOpen
  \bibfield  {author} {\bibinfo {author} {\bibfnamefont {M.}~\bibnamefont
  {Vergniory}}, \bibinfo {author} {\bibfnamefont {L.}~\bibnamefont {Elcoro}},
  \bibinfo {author} {\bibfnamefont {C.}~\bibnamefont {Felser}}, \bibinfo
  {author} {\bibfnamefont {B.}~\bibnamefont {Bernevig}}, \ and\ \bibinfo
  {author} {\bibfnamefont {Z.}~\bibnamefont {Wang}},\ }\href@noop {} {\bibfield
   {journal} {\bibinfo  {journal} {arXiv preprint arXiv:1807.10271}\ }
  (\bibinfo {year} {2018})}\BibitemShut {NoStop}%
\bibitem [{\citenamefont {Agarwala}\ and\ \citenamefont
  {Shenoy}(2017)}]{Agarwala_PRL_2017}%
  \BibitemOpen
  \bibfield  {author} {\bibinfo {author} {\bibfnamefont {A.}~\bibnamefont
  {Agarwala}}\ and\ \bibinfo {author} {\bibfnamefont {V.~B.}\ \bibnamefont
  {Shenoy}},\ }\href {\doibase 10.1103/PhysRevLett.118.236402} {\bibfield
  {journal} {\bibinfo  {journal} {Phys. Rev. Lett.}\ }\textbf {\bibinfo
  {volume} {118}},\ \bibinfo {pages} {236402} (\bibinfo {year}
  {2017})}\BibitemShut {NoStop}%
\bibitem [{\citenamefont {Mitchell}\ \emph {et~al.}(2018)\citenamefont
  {Mitchell}, \citenamefont {Nash}, \citenamefont {Hexner}, \citenamefont
  {Turner},\ and\ \citenamefont {Irvine}}]{Mitchell_NP_2018}%
  \BibitemOpen
  \bibfield  {author} {\bibinfo {author} {\bibfnamefont {N.~P.}\ \bibnamefont
  {Mitchell}}, \bibinfo {author} {\bibfnamefont {L.~M.}\ \bibnamefont {Nash}},
  \bibinfo {author} {\bibfnamefont {D.}~\bibnamefont {Hexner}}, \bibinfo
  {author} {\bibfnamefont {A.~M.}\ \bibnamefont {Turner}}, \ and\ \bibinfo
  {author} {\bibfnamefont {W.~T.}\ \bibnamefont {Irvine}},\ }\href@noop {}
  {\bibfield  {journal} {\bibinfo  {journal} {Nature Physics}\ ,\ \bibinfo
  {pages} {1}} (\bibinfo {year} {2018})}\BibitemShut {NoStop}%
\bibitem [{\citenamefont {Shindou}\ and\ \citenamefont
  {Murakami}(2009)}]{Shindou_PRB_2009}%
  \BibitemOpen
  \bibfield  {author} {\bibinfo {author} {\bibfnamefont {R.}~\bibnamefont
  {Shindou}}\ and\ \bibinfo {author} {\bibfnamefont {S.}~\bibnamefont
  {Murakami}},\ }\href {\doibase 10.1103/PhysRevB.79.045321} {\bibfield
  {journal} {\bibinfo  {journal} {Phys. Rev. B}\ }\textbf {\bibinfo {volume}
  {79}},\ \bibinfo {pages} {045321} (\bibinfo {year} {2009})}\BibitemShut
  {NoStop}%
\bibitem [{\citenamefont {Leung}\ and\ \citenamefont
  {Prodan}(2012)}]{Leung_PRB_2012}%
  \BibitemOpen
  \bibfield  {author} {\bibinfo {author} {\bibfnamefont {B.}~\bibnamefont
  {Leung}}\ and\ \bibinfo {author} {\bibfnamefont {E.}~\bibnamefont {Prodan}},\
  }\href {\doibase 10.1103/PhysRevB.85.205136} {\bibfield  {journal} {\bibinfo
  {journal} {Phys. Rev. B}\ }\textbf {\bibinfo {volume} {85}},\ \bibinfo
  {pages} {205136} (\bibinfo {year} {2012})}\BibitemShut {NoStop}%
\bibitem [{\citenamefont {Kobayashi}\ \emph {et~al.}(2013)\citenamefont
  {Kobayashi}, \citenamefont {Ohtsuki},\ and\ \citenamefont
  {Imura}}]{Kobayashi_PRL_2013}%
  \BibitemOpen
  \bibfield  {author} {\bibinfo {author} {\bibfnamefont {K.}~\bibnamefont
  {Kobayashi}}, \bibinfo {author} {\bibfnamefont {T.}~\bibnamefont {Ohtsuki}},
  \ and\ \bibinfo {author} {\bibfnamefont {K.-I.}\ \bibnamefont {Imura}},\
  }\href {\doibase 10.1103/PhysRevLett.110.236803} {\bibfield  {journal}
  {\bibinfo  {journal} {Phys. Rev. Lett.}\ }\textbf {\bibinfo {volume} {110}},\
  \bibinfo {pages} {236803} (\bibinfo {year} {2013})}\BibitemShut {NoStop}%
\bibitem [{\citenamefont {Song}\ \emph {et~al.}(2014)\citenamefont {Song},
  \citenamefont {Fine},\ and\ \citenamefont {Prodan}}]{Song_PRB_2014}%
  \BibitemOpen
  \bibfield  {author} {\bibinfo {author} {\bibfnamefont {J.}~\bibnamefont
  {Song}}, \bibinfo {author} {\bibfnamefont {C.}~\bibnamefont {Fine}}, \ and\
  \bibinfo {author} {\bibfnamefont {E.}~\bibnamefont {Prodan}},\ }\href
  {\doibase 10.1103/PhysRevB.90.184201} {\bibfield  {journal} {\bibinfo
  {journal} {Phys. Rev. B}\ }\textbf {\bibinfo {volume} {90}},\ \bibinfo
  {pages} {184201} (\bibinfo {year} {2014})}\BibitemShut {NoStop}%
\bibitem [{\citenamefont {Ryu}\ and\ \citenamefont
  {Nomura}(2012)}]{Ryu_PRB_2012}%
  \BibitemOpen
  \bibfield  {author} {\bibinfo {author} {\bibfnamefont {S.}~\bibnamefont
  {Ryu}}\ and\ \bibinfo {author} {\bibfnamefont {K.}~\bibnamefont {Nomura}},\
  }\href {\doibase 10.1103/PhysRevB.85.155138} {\bibfield  {journal} {\bibinfo
  {journal} {Phys. Rev. B}\ }\textbf {\bibinfo {volume} {85}},\ \bibinfo
  {pages} {155138} (\bibinfo {year} {2012})}\BibitemShut {NoStop}%
\bibitem [{\citenamefont {Sbierski}\ and\ \citenamefont
  {Brouwer}(2014)}]{Sbierski_PRB_2014}%
  \BibitemOpen
  \bibfield  {author} {\bibinfo {author} {\bibfnamefont {B.}~\bibnamefont
  {Sbierski}}\ and\ \bibinfo {author} {\bibfnamefont {P.~W.}\ \bibnamefont
  {Brouwer}},\ }\href {\doibase 10.1103/PhysRevB.89.155311} {\bibfield
  {journal} {\bibinfo  {journal} {Phys. Rev. B}\ }\textbf {\bibinfo {volume}
  {89}},\ \bibinfo {pages} {155311} (\bibinfo {year} {2014})}\BibitemShut
  {NoStop}%
\bibitem [{\citenamefont {Ohtsuki}\ and\ \citenamefont
  {Ohtsuki}(2017)}]{Ohtsuki_JPSJ_2017}%
  \BibitemOpen
  \bibfield  {author} {\bibinfo {author} {\bibfnamefont {T.}~\bibnamefont
  {Ohtsuki}}\ and\ \bibinfo {author} {\bibfnamefont {T.}~\bibnamefont
  {Ohtsuki}},\ }\href {\doibase 10.7566/JPSJ.86.044708} {\bibfield  {journal}
  {\bibinfo  {journal} {Journal of the Physical Society of Japan}\ }\textbf
  {\bibinfo {volume} {86}},\ \bibinfo {pages} {044708} (\bibinfo {year}
  {2017})},\ \Eprint
  {http://arxiv.org/abs/https://doi.org/10.7566/JPSJ.86.04470}
  {https://doi.org/10.7566/JPSJ.86.04470} \BibitemShut {NoStop}%
\bibitem [{\citenamefont {Akagi}\ \emph {et~al.}(2017)\citenamefont {Akagi},
  \citenamefont {Katsura},\ and\ \citenamefont {Koma}}]{Akagi_JPSJ_2017}%
  \BibitemOpen
  \bibfield  {author} {\bibinfo {author} {\bibfnamefont {Y.}~\bibnamefont
  {Akagi}}, \bibinfo {author} {\bibfnamefont {H.}~\bibnamefont {Katsura}}, \
  and\ \bibinfo {author} {\bibfnamefont {T.}~\bibnamefont {Koma}},\ }\href
  {\doibase 10.7566/JPSJ.86.123710} {\bibfield  {journal} {\bibinfo  {journal}
  {Journal of the Physical Society of Japan}\ }\textbf {\bibinfo {volume}
  {86}},\ \bibinfo {pages} {123710} (\bibinfo {year} {2017})},\ \Eprint
  {http://arxiv.org/abs/https://doi.org/10.7566/JPSJ.86.123710}
  {https://doi.org/10.7566/JPSJ.86.123710} \BibitemShut {NoStop}%
\bibitem [{\citenamefont {Guo}\ \emph {et~al.}(2010)\citenamefont {Guo},
  \citenamefont {Rosenberg}, \citenamefont {Refael},\ and\ \citenamefont
  {Franz}}]{Guo_PRL_2010}%
  \BibitemOpen
  \bibfield  {author} {\bibinfo {author} {\bibfnamefont {H.-M.}\ \bibnamefont
  {Guo}}, \bibinfo {author} {\bibfnamefont {G.}~\bibnamefont {Rosenberg}},
  \bibinfo {author} {\bibfnamefont {G.}~\bibnamefont {Refael}}, \ and\ \bibinfo
  {author} {\bibfnamefont {M.}~\bibnamefont {Franz}},\ }\href {\doibase
  10.1103/PhysRevLett.105.216601} {\bibfield  {journal} {\bibinfo  {journal}
  {Phys. Rev. Lett.}\ }\textbf {\bibinfo {volume} {105}},\ \bibinfo {pages}
  {216601} (\bibinfo {year} {2010})}\BibitemShut {NoStop}%
\bibitem [{\citenamefont {Prodan}(2017)}]{Prodan_2017}%
  \BibitemOpen
  \bibfield  {author} {\bibinfo {author} {\bibfnamefont {E.}~\bibnamefont
  {Prodan}},\ }\enquote {\bibinfo {title} {Disordered topological insulators: A
  brief introduction},}\ in\ \href {\doibase 10.1007/978-3-319-55023-7_1}
  {\emph {\bibinfo {booktitle} {A Computational Non-commutative Geometry
  Program for Disordered Topological Insulators}}}\ (\bibinfo  {publisher}
  {Springer International Publishing},\ \bibinfo {address} {Cham},\ \bibinfo
  {year} {2017})\ pp.\ \bibinfo {pages} {1--9}\BibitemShut {NoStop}%
\bibitem [{\citenamefont {Huang}\ and\ \citenamefont
  {Liu}(2018)}]{Huang_PRL_2018}%
  \BibitemOpen
  \bibfield  {author} {\bibinfo {author} {\bibfnamefont {H.}~\bibnamefont
  {Huang}}\ and\ \bibinfo {author} {\bibfnamefont {F.}~\bibnamefont {Liu}},\
  }\href {\doibase 10.1103/PhysRevLett.121.126401} {\bibfield  {journal}
  {\bibinfo  {journal} {Phys. Rev. Lett.}\ }\textbf {\bibinfo {volume} {121}},\
  \bibinfo {pages} {126401} (\bibinfo {year} {2018})}\BibitemShut {NoStop}%
\bibitem [{\citenamefont {{Li}}\ and\ \citenamefont
  {{Mong}}(2019)}]{Li_arXiv_2019}%
  \BibitemOpen
  \bibfield  {author} {\bibinfo {author} {\bibfnamefont {Z.}~\bibnamefont
  {{Li}}}\ and\ \bibinfo {author} {\bibfnamefont {R.~S.~K.}\ \bibnamefont
  {{Mong}}},\ }\href@noop {} {\bibfield  {journal} {\bibinfo  {journal} {arXiv
  e-prints}\ ,\ \bibinfo {eid} {arXiv:1905.12649}} (\bibinfo {year} {2019})},\
  \Eprint {http://arxiv.org/abs/1905.12649} {arXiv:1905.12649
  [cond-mat.mes-hall]} \BibitemShut {NoStop}%
\bibitem [{\citenamefont {Bourne}\ \emph {et~al.}(2016)\citenamefont {Bourne},
  \citenamefont {Carey},\ and\ \citenamefont {Rennie}}]{Bourne_RMP_2016}%
  \BibitemOpen
  \bibfield  {author} {\bibinfo {author} {\bibfnamefont {C.}~\bibnamefont
  {Bourne}}, \bibinfo {author} {\bibfnamefont {A.~L.}\ \bibnamefont {Carey}}, \
  and\ \bibinfo {author} {\bibfnamefont {A.}~\bibnamefont {Rennie}},\
  }\href@noop {} {\bibfield  {journal} {\bibinfo  {journal} {Reviews in
  Mathematical Physics}\ }\textbf {\bibinfo {volume} {28}},\ \bibinfo {pages}
  {1650004} (\bibinfo {year} {2016})}\BibitemShut {NoStop}%
\bibitem [{\citenamefont {Loring}(2015)}]{Loring_AOP_2015}%
  \BibitemOpen
  \bibfield  {author} {\bibinfo {author} {\bibfnamefont {T.~A.}\ \bibnamefont
  {Loring}},\ }\href {\doibase https://doi.org/10.1016/j.aop.2015.02.031}
  {\bibfield  {journal} {\bibinfo  {journal} {Annals of Physics}\ }\textbf
  {\bibinfo {volume} {356}},\ \bibinfo {pages} {383 } (\bibinfo {year}
  {2015})}\BibitemShut {NoStop}%
\bibitem [{\citenamefont {Prodan}\ and\ \citenamefont
  {Schulz-Baldes}(2016{\natexlab{a}})}]{Prodan2016bulk}%
  \BibitemOpen
  \bibfield  {author} {\bibinfo {author} {\bibfnamefont {E.}~\bibnamefont
  {Prodan}}\ and\ \bibinfo {author} {\bibfnamefont {H.}~\bibnamefont
  {Schulz-Baldes}},\ }\href@noop {} {\bibfield  {journal} {\bibinfo  {journal}
  {K}\ } (\bibinfo {year} {2016}{\natexlab{a}})}\BibitemShut {NoStop}%
\bibitem [{\citenamefont {Essin}\ and\ \citenamefont
  {Moore}(2007)}]{Essin_PRB_2007}%
  \BibitemOpen
  \bibfield  {author} {\bibinfo {author} {\bibfnamefont {A.~M.}\ \bibnamefont
  {Essin}}\ and\ \bibinfo {author} {\bibfnamefont {J.~E.}\ \bibnamefont
  {Moore}},\ }\href {\doibase 10.1103/PhysRevB.76.165307} {\bibfield  {journal}
  {\bibinfo  {journal} {Phys. Rev. B}\ }\textbf {\bibinfo {volume} {76}},\
  \bibinfo {pages} {165307} (\bibinfo {year} {2007})}\BibitemShut {NoStop}%
\bibitem [{\citenamefont {Prodan}\ and\ \citenamefont
  {Schulz-Baldes}(2016{\natexlab{b}})}]{Prodan_JFA_2016}%
  \BibitemOpen
  \bibfield  {author} {\bibinfo {author} {\bibfnamefont {E.}~\bibnamefont
  {Prodan}}\ and\ \bibinfo {author} {\bibfnamefont {H.}~\bibnamefont
  {Schulz-Baldes}},\ }\href {\doibase
  https://doi.org/10.1016/j.jfa.2016.06.001} {\bibfield  {journal} {\bibinfo
  {journal} {Journal of Functional Analysis}\ }\textbf {\bibinfo {volume}
  {271}},\ \bibinfo {pages} {1150 } (\bibinfo {year}
  {2016}{\natexlab{b}})}\BibitemShut {NoStop}%
\bibitem [{\citenamefont {Li}\ \emph {et~al.}(2009)\citenamefont {Li},
  \citenamefont {Chu}, \citenamefont {Jain},\ and\ \citenamefont
  {Shen}}]{Jian_PRL_2009}%
  \BibitemOpen
  \bibfield  {author} {\bibinfo {author} {\bibfnamefont {J.}~\bibnamefont
  {Li}}, \bibinfo {author} {\bibfnamefont {R.-L.}\ \bibnamefont {Chu}},
  \bibinfo {author} {\bibfnamefont {J.~K.}\ \bibnamefont {Jain}}, \ and\
  \bibinfo {author} {\bibfnamefont {S.-Q.}\ \bibnamefont {Shen}},\ }\href
  {\doibase 10.1103/PhysRevLett.102.136806} {\bibfield  {journal} {\bibinfo
  {journal} {Phys. Rev. Lett.}\ }\textbf {\bibinfo {volume} {102}},\ \bibinfo
  {pages} {136806} (\bibinfo {year} {2009})}\BibitemShut {NoStop}%
\bibitem [{\citenamefont {Loring}\ and\ \citenamefont
  {Hastings}(2010)}]{Loring_EPL_2010}%
  \BibitemOpen
  \bibfield  {author} {\bibinfo {author} {\bibfnamefont {T.~A.}\ \bibnamefont
  {Loring}}\ and\ \bibinfo {author} {\bibfnamefont {M.~B.}\ \bibnamefont
  {Hastings}},\ }\href {\doibase 10.1209/0295-5075/92/67004} {\bibfield
  {journal} {\bibinfo  {journal} {{EPL} (Europhysics Letters)}\ }\textbf
  {\bibinfo {volume} {92}},\ \bibinfo {pages} {67004} (\bibinfo {year}
  {2010})}\BibitemShut {NoStop}%
\bibitem [{\citenamefont {Wang}\ and\ \citenamefont
  {Zhang}(2014)}]{Wang_PRX_2014}%
  \BibitemOpen
  \bibfield  {author} {\bibinfo {author} {\bibfnamefont {Z.}~\bibnamefont
  {Wang}}\ and\ \bibinfo {author} {\bibfnamefont {S.-C.}\ \bibnamefont
  {Zhang}},\ }\href {\doibase 10.1103/PhysRevX.4.011006} {\bibfield  {journal}
  {\bibinfo  {journal} {Phys. Rev. X}\ }\textbf {\bibinfo {volume} {4}},\
  \bibinfo {pages} {011006} (\bibinfo {year} {2014})}\BibitemShut {NoStop}%
\bibitem [{\citenamefont {Guo}(2010)}]{Guo_PRB_2010}%
  \BibitemOpen
  \bibfield  {author} {\bibinfo {author} {\bibfnamefont {H.-M.}\ \bibnamefont
  {Guo}},\ }\href {\doibase 10.1103/PhysRevB.82.115122} {\bibfield  {journal}
  {\bibinfo  {journal} {Phys. Rev. B}\ }\textbf {\bibinfo {volume} {82}},\
  \bibinfo {pages} {115122} (\bibinfo {year} {2010})}\BibitemShut {NoStop}%
\bibitem [{\citenamefont {{Varjas}}\ \emph {et~al.}(2019)\citenamefont
  {{Varjas}}, \citenamefont {{Fruchart}}, \citenamefont {{Akhmerov}},\ and\
  \citenamefont {{Perez-Piskunow}}}]{Varjas_arXiv_2019}%
  \BibitemOpen
  \bibfield  {author} {\bibinfo {author} {\bibfnamefont {D.}~\bibnamefont
  {{Varjas}}}, \bibinfo {author} {\bibfnamefont {M.}~\bibnamefont
  {{Fruchart}}}, \bibinfo {author} {\bibfnamefont {A.~R.}\ \bibnamefont
  {{Akhmerov}}}, \ and\ \bibinfo {author} {\bibfnamefont {P.}~\bibnamefont
  {{Perez-Piskunow}}},\ }\href@noop {} {\bibfield  {journal} {\bibinfo
  {journal} {arXiv e-prints}\ ,\ \bibinfo {eid} {arXiv:1905.02215}} (\bibinfo
  {year} {2019})},\ \Eprint {http://arxiv.org/abs/1905.02215} {arXiv:1905.02215
  [cond-mat.mes-hall]} \BibitemShut {NoStop}%
\bibitem [{\citenamefont {{Mondragon-Shem}}\ and\ \citenamefont
  {{Hughes}}(2019)}]{Shem_arXiv_2019}%
  \BibitemOpen
  \bibfield  {author} {\bibinfo {author} {\bibfnamefont {I.}~\bibnamefont
  {{Mondragon-Shem}}}\ and\ \bibinfo {author} {\bibfnamefont {T.~L.}\
  \bibnamefont {{Hughes}}},\ }\href@noop {} {\bibfield  {journal} {\bibinfo
  {journal} {arXiv e-prints}\ ,\ \bibinfo {eid} {arXiv:1906.11847}} (\bibinfo
  {year} {2019})},\ \Eprint {http://arxiv.org/abs/1906.11847} {arXiv:1906.11847
  [cond-mat.dis-nn]} \BibitemShut {NoStop}%
\bibitem [{\citenamefont {Yamakage}\ \emph {et~al.}(2013)\citenamefont
  {Yamakage}, \citenamefont {Nomura}, \citenamefont {Imura},\ and\
  \citenamefont {Kuramoto}}]{Yamakage_PRB_2013}%
  \BibitemOpen
  \bibfield  {author} {\bibinfo {author} {\bibfnamefont {A.}~\bibnamefont
  {Yamakage}}, \bibinfo {author} {\bibfnamefont {K.}~\bibnamefont {Nomura}},
  \bibinfo {author} {\bibfnamefont {K.-I.}\ \bibnamefont {Imura}}, \ and\
  \bibinfo {author} {\bibfnamefont {Y.}~\bibnamefont {Kuramoto}},\ }\href
  {\doibase 10.1103/PhysRevB.87.205141} {\bibfield  {journal} {\bibinfo
  {journal} {Phys. Rev. B}\ }\textbf {\bibinfo {volume} {87}},\ \bibinfo
  {pages} {205141} (\bibinfo {year} {2013})}\BibitemShut {NoStop}%
\bibitem [{\citenamefont {Banerjee}\ \emph {et~al.}(2017)\citenamefont
  {Banerjee}, \citenamefont {Deb}, \citenamefont {Majhi}, \citenamefont
  {Ganesan}, \citenamefont {Sen},\ and\ \citenamefont
  {Anil~Kumar}}]{Banerjee_NS_2017}%
  \BibitemOpen
  \bibfield  {author} {\bibinfo {author} {\bibfnamefont {A.}~\bibnamefont
  {Banerjee}}, \bibinfo {author} {\bibfnamefont {O.}~\bibnamefont {Deb}},
  \bibinfo {author} {\bibfnamefont {K.}~\bibnamefont {Majhi}}, \bibinfo
  {author} {\bibfnamefont {R.}~\bibnamefont {Ganesan}}, \bibinfo {author}
  {\bibfnamefont {D.}~\bibnamefont {Sen}}, \ and\ \bibinfo {author}
  {\bibfnamefont {P.~S.}\ \bibnamefont {Anil~Kumar}},\ }\href {\doibase
  10.1039/C7NR01355H} {\bibfield  {journal} {\bibinfo  {journal} {Nanoscale}\
  }\textbf {\bibinfo {volume} {9}},\ \bibinfo {pages} {6755} (\bibinfo {year}
  {2017})}\BibitemShut {NoStop}%
\bibitem [{\citenamefont {Roy}\ \emph {et~al.}(2017)\citenamefont {Roy},
  \citenamefont {Alavirad},\ and\ \citenamefont {Sau}}]{Roy_PRL_2017}%
  \BibitemOpen
  \bibfield  {author} {\bibinfo {author} {\bibfnamefont {B.}~\bibnamefont
  {Roy}}, \bibinfo {author} {\bibfnamefont {Y.}~\bibnamefont {Alavirad}}, \
  and\ \bibinfo {author} {\bibfnamefont {J.~D.}\ \bibnamefont {Sau}},\ }\href
  {\doibase 10.1103/PhysRevLett.118.227002} {\bibfield  {journal} {\bibinfo
  {journal} {Phys. Rev. Lett.}\ }\textbf {\bibinfo {volume} {118}},\ \bibinfo
  {pages} {227002} (\bibinfo {year} {2017})}\BibitemShut {NoStop}%
\bibitem [{\citenamefont {Goswami}\ and\ \citenamefont
  {Chakravarty}(2017)}]{Goswami_PRB_2017}%
  \BibitemOpen
  \bibfield  {author} {\bibinfo {author} {\bibfnamefont {P.}~\bibnamefont
  {Goswami}}\ and\ \bibinfo {author} {\bibfnamefont {S.}~\bibnamefont
  {Chakravarty}},\ }\href {\doibase 10.1103/PhysRevB.95.075131} {\bibfield
  {journal} {\bibinfo  {journal} {Phys. Rev. B}\ }\textbf {\bibinfo {volume}
  {95}},\ \bibinfo {pages} {075131} (\bibinfo {year} {2017})}\BibitemShut
  {NoStop}%
\bibitem [{\citenamefont {Liu}\ \emph {et~al.}(2017)\citenamefont {Liu},
  \citenamefont {Gao}, \citenamefont {Yang},\ and\ \citenamefont
  {Zhang}}]{Liu_PRL_2017}%
  \BibitemOpen
  \bibfield  {author} {\bibinfo {author} {\bibfnamefont {C.}~\bibnamefont
  {Liu}}, \bibinfo {author} {\bibfnamefont {W.}~\bibnamefont {Gao}}, \bibinfo
  {author} {\bibfnamefont {B.}~\bibnamefont {Yang}}, \ and\ \bibinfo {author}
  {\bibfnamefont {S.}~\bibnamefont {Zhang}},\ }\href {\doibase
  10.1103/PhysRevLett.119.183901} {\bibfield  {journal} {\bibinfo  {journal}
  {Phys. Rev. Lett.}\ }\textbf {\bibinfo {volume} {119}},\ \bibinfo {pages}
  {183901} (\bibinfo {year} {2017})}\BibitemShut {NoStop}%
\bibitem [{\citenamefont {{Loring}}(2018)}]{Loring_arXiv_2018}%
  \BibitemOpen
  \bibfield  {author} {\bibinfo {author} {\bibfnamefont {T.~A.}\ \bibnamefont
  {{Loring}}},\ }\href@noop {} {\bibfield  {journal} {\bibinfo  {journal}
  {arXiv e-prints}\ ,\ \bibinfo {eid} {arXiv:1811.07494}} (\bibinfo {year}
  {2018})},\ \Eprint {http://arxiv.org/abs/1811.07494} {arXiv:1811.07494
  [cond-mat.mes-hall]} \BibitemShut {NoStop}%
\bibitem [{\citenamefont {Chern}(2019)}]{Chern_EPL_2019}%
  \BibitemOpen
  \bibfield  {author} {\bibinfo {author} {\bibfnamefont {G.-W.}\ \bibnamefont
  {Chern}},\ }\href {\doibase 10.1209/0295-5075/126/37002} {\bibfield
  {journal} {\bibinfo  {journal} {{EPL} (Europhysics Letters)}\ }\textbf
  {\bibinfo {volume} {126}},\ \bibinfo {pages} {37002} (\bibinfo {year}
  {2019})}\BibitemShut {NoStop}%
\bibitem [{\citenamefont {Peano}\ and\ \citenamefont
  {Schulz-Baldes}(2018)}]{Vittorio_JMP_2018}%
  \BibitemOpen
  \bibfield  {author} {\bibinfo {author} {\bibfnamefont {V.}~\bibnamefont
  {Peano}}\ and\ \bibinfo {author} {\bibfnamefont {H.}~\bibnamefont
  {Schulz-Baldes}},\ }\href {\doibase 10.1063/1.5002094} {\bibfield  {journal}
  {\bibinfo  {journal} {Journal of Mathematical Physics}\ }\textbf {\bibinfo
  {volume} {59}},\ \bibinfo {pages} {031901} (\bibinfo {year} {2018})},\
  \Eprint {http://arxiv.org/abs/https://doi.org/10.1063/1.5002094}
  {https://doi.org/10.1063/1.5002094} \BibitemShut {NoStop}%
\bibitem [{\citenamefont {Lee}\ and\ \citenamefont
  {Ramakrishnan}(1985)}]{Lee_RMP_1985}%
  \BibitemOpen
  \bibfield  {author} {\bibinfo {author} {\bibfnamefont {P.~A.}\ \bibnamefont
  {Lee}}\ and\ \bibinfo {author} {\bibfnamefont {T.~V.}\ \bibnamefont
  {Ramakrishnan}},\ }\href {\doibase 10.1103/RevModPhys.57.287} {\bibfield
  {journal} {\bibinfo  {journal} {Rev. Mod. Phys.}\ }\textbf {\bibinfo {volume}
  {57}},\ \bibinfo {pages} {287} (\bibinfo {year} {1985})}\BibitemShut
  {NoStop}%
\bibitem [{\citenamefont {Edwards}\ and\ \citenamefont
  {Thouless}(1972)}]{Edwards_1972}%
  \BibitemOpen
  \bibfield  {author} {\bibinfo {author} {\bibfnamefont {J.~T.}\ \bibnamefont
  {Edwards}}\ and\ \bibinfo {author} {\bibfnamefont {D.~J.}\ \bibnamefont
  {Thouless}},\ }\href {\doibase 10.1088/0022-3719/5/8/007} {\bibfield
  {journal} {\bibinfo  {journal} {Journal of Physics C: Solid State Physics}\
  }\textbf {\bibinfo {volume} {5}},\ \bibinfo {pages} {807} (\bibinfo {year}
  {1972})}\BibitemShut {NoStop}%
\bibitem [{\citenamefont {Thouless}(1974)}]{Thouless_1974}%
  \BibitemOpen
  \bibfield  {author} {\bibinfo {author} {\bibfnamefont {D.}~\bibnamefont
  {Thouless}},\ }\href {\doibase https://doi.org/10.1016/0370-1573(74)90029-5}
  {\bibfield  {journal} {\bibinfo  {journal} {Physics Reports}\ }\textbf
  {\bibinfo {volume} {13}},\ \bibinfo {pages} {93 } (\bibinfo {year}
  {1974})}\BibitemShut {NoStop}%
\bibitem [{\citenamefont {Kramer}\ and\ \citenamefont
  {MacKinnon}(1993)}]{Kramer_RPP_1993}%
  \BibitemOpen
  \bibfield  {author} {\bibinfo {author} {\bibfnamefont {B.}~\bibnamefont
  {Kramer}}\ and\ \bibinfo {author} {\bibfnamefont {A.}~\bibnamefont
  {MacKinnon}},\ }\href {\doibase 10.1088/0034-4885/56/12/001} {\bibfield
  {journal} {\bibinfo  {journal} {Reports on Progress in Physics}\ }\textbf
  {\bibinfo {volume} {56}},\ \bibinfo {pages} {1469} (\bibinfo {year}
  {1993})}\BibitemShut {NoStop}%
\bibitem [{\citenamefont {Evers}\ and\ \citenamefont
  {Mirlin}(2008)}]{Evers_RMP_2008}%
  \BibitemOpen
  \bibfield  {author} {\bibinfo {author} {\bibfnamefont {F.}~\bibnamefont
  {Evers}}\ and\ \bibinfo {author} {\bibfnamefont {A.~D.}\ \bibnamefont
  {Mirlin}},\ }\href {\doibase 10.1103/RevModPhys.80.1355} {\bibfield
  {journal} {\bibinfo  {journal} {Rev. Mod. Phys.}\ }\textbf {\bibinfo {volume}
  {80}},\ \bibinfo {pages} {1355} (\bibinfo {year} {2008})}\BibitemShut
  {NoStop}%
\bibitem [{\citenamefont {{Sahlberg}}\ \emph {et~al.}(2019)\citenamefont
  {{Sahlberg}}, \citenamefont {{Weststr{\"o}m}}, \citenamefont
  {{P{\"o}yh{\"o}nen}},\ and\ \citenamefont {{Ojanen}}}]{Sahlberg_arXiv_2019}%
  \BibitemOpen
  \bibfield  {author} {\bibinfo {author} {\bibfnamefont {I.}~\bibnamefont
  {{Sahlberg}}}, \bibinfo {author} {\bibfnamefont {A.}~\bibnamefont
  {{Weststr{\"o}m}}}, \bibinfo {author} {\bibfnamefont {K.}~\bibnamefont
  {{P{\"o}yh{\"o}nen}}}, \ and\ \bibinfo {author} {\bibfnamefont
  {T.}~\bibnamefont {{Ojanen}}},\ }\href@noop {} {\bibfield  {journal}
  {\bibinfo  {journal} {arXiv e-prints}\ ,\ \bibinfo {eid} {arXiv:1902.01623}}
  (\bibinfo {year} {2019})},\ \Eprint {http://arxiv.org/abs/1902.01623}
  {arXiv:1902.01623 [cond-mat.mes-hall]} \BibitemShut {NoStop}%
\bibitem [{\citenamefont {Witten}(1979)}]{Witten_PLB_1979}%
  \BibitemOpen
  \bibfield  {author} {\bibinfo {author} {\bibfnamefont {E.}~\bibnamefont
  {Witten}},\ }\href {\doibase https://doi.org/10.1016/0370-2693(79)90838-4}
  {\bibfield  {journal} {\bibinfo  {journal} {Physics Letters B}\ }\textbf
  {\bibinfo {volume} {86}},\ \bibinfo {pages} {283 } (\bibinfo {year}
  {1979})}\BibitemShut {NoStop}%
\bibitem [{\citenamefont {Wilczek}(1987)}]{Wilczek_PRL_1987}%
  \BibitemOpen
  \bibfield  {author} {\bibinfo {author} {\bibfnamefont {F.}~\bibnamefont
  {Wilczek}},\ }\href {\doibase 10.1103/PhysRevLett.58.1799} {\bibfield
  {journal} {\bibinfo  {journal} {Phys. Rev. Lett.}\ }\textbf {\bibinfo
  {volume} {58}},\ \bibinfo {pages} {1799} (\bibinfo {year}
  {1987})}\BibitemShut {NoStop}%
\bibitem [{\citenamefont {Rosenberg}\ and\ \citenamefont
  {Franz}(2010)}]{Rosenberg_PRB_2010}%
  \BibitemOpen
  \bibfield  {author} {\bibinfo {author} {\bibfnamefont {G.}~\bibnamefont
  {Rosenberg}}\ and\ \bibinfo {author} {\bibfnamefont {M.}~\bibnamefont
  {Franz}},\ }\href {\doibase 10.1103/PhysRevB.82.035105} {\bibfield  {journal}
  {\bibinfo  {journal} {Phys. Rev. B}\ }\textbf {\bibinfo {volume} {82}},\
  \bibinfo {pages} {035105} (\bibinfo {year} {2010})}\BibitemShut {NoStop}%
\bibitem [{\citenamefont {Heine}(1971)}]{Heine_JPC_1971}%
  \BibitemOpen
  \bibfield  {author} {\bibinfo {author} {\bibfnamefont {V.}~\bibnamefont
  {Heine}},\ }\href {http://stacks.iop.org/0022-3719/4/i=10/a=012} {\bibfield
  {journal} {\bibinfo  {journal} {Journal of Physics C: Solid State Physics}\
  }\textbf {\bibinfo {volume} {4}},\ \bibinfo {pages} {L221} (\bibinfo {year}
  {1971})}\BibitemShut {NoStop}%
\bibitem [{\citenamefont {Weaire}\ and\ \citenamefont
  {Thorpe}(1971)}]{Wearie_PRB_1971}%
  \BibitemOpen
  \bibfield  {author} {\bibinfo {author} {\bibfnamefont {D.}~\bibnamefont
  {Weaire}}\ and\ \bibinfo {author} {\bibfnamefont {M.~F.}\ \bibnamefont
  {Thorpe}},\ }\href {\doibase 10.1103/PhysRevB.4.2508} {\bibfield  {journal}
  {\bibinfo  {journal} {Phys. Rev. B}\ }\textbf {\bibinfo {volume} {4}},\
  \bibinfo {pages} {2508} (\bibinfo {year} {1971})}\BibitemShut {NoStop}%
\bibitem [{\citenamefont {Thorpe}\ and\ \citenamefont
  {Weaire}(1971)}]{Thorpe_PRB_1971}%
  \BibitemOpen
  \bibfield  {author} {\bibinfo {author} {\bibfnamefont {M.~F.}\ \bibnamefont
  {Thorpe}}\ and\ \bibinfo {author} {\bibfnamefont {D.}~\bibnamefont
  {Weaire}},\ }\href {\doibase 10.1103/PhysRevB.4.3518} {\bibfield  {journal}
  {\bibinfo  {journal} {Phys. Rev. B}\ }\textbf {\bibinfo {volume} {4}},\
  \bibinfo {pages} {3518} (\bibinfo {year} {1971})}\BibitemShut {NoStop}%
\bibitem [{\citenamefont {Elliott}\ \emph {et~al.}(1974)\citenamefont
  {Elliott}, \citenamefont {Krumhansl},\ and\ \citenamefont
  {Leath}}]{Elliott_RMP_1974}%
  \BibitemOpen
  \bibfield  {author} {\bibinfo {author} {\bibfnamefont {R.~J.}\ \bibnamefont
  {Elliott}}, \bibinfo {author} {\bibfnamefont {J.~A.}\ \bibnamefont
  {Krumhansl}}, \ and\ \bibinfo {author} {\bibfnamefont {P.~L.}\ \bibnamefont
  {Leath}},\ }\href {\doibase 10.1103/RevModPhys.46.465} {\bibfield  {journal}
  {\bibinfo  {journal} {Rev. Mod. Phys.}\ }\textbf {\bibinfo {volume} {46}},\
  \bibinfo {pages} {465} (\bibinfo {year} {1974})}\BibitemShut {NoStop}%
\bibitem [{\citenamefont {Schwartz}\ and\ \citenamefont
  {Ehrenreich}(1972)}]{Schwartz_PRB_1972}%
  \BibitemOpen
  \bibfield  {author} {\bibinfo {author} {\bibfnamefont {L.}~\bibnamefont
  {Schwartz}}\ and\ \bibinfo {author} {\bibfnamefont {H.}~\bibnamefont
  {Ehrenreich}},\ }\href {\doibase 10.1103/PhysRevB.6.4088} {\bibfield
  {journal} {\bibinfo  {journal} {Phys. Rev. B}\ }\textbf {\bibinfo {volume}
  {6}},\ \bibinfo {pages} {4088} (\bibinfo {year} {1972})}\BibitemShut
  {NoStop}%
\bibitem [{\citenamefont {Davies}\ \emph {et~al.}(1984)\citenamefont {Davies},
  \citenamefont {Lee},\ and\ \citenamefont {Rice}}]{Davies_PRB_1984}%
  \BibitemOpen
  \bibfield  {author} {\bibinfo {author} {\bibfnamefont {J.~H.}\ \bibnamefont
  {Davies}}, \bibinfo {author} {\bibfnamefont {P.~A.}\ \bibnamefont {Lee}}, \
  and\ \bibinfo {author} {\bibfnamefont {T.~M.}\ \bibnamefont {Rice}},\ }\href
  {\doibase 10.1103/PhysRevB.29.4260} {\bibfield  {journal} {\bibinfo
  {journal} {Phys. Rev. B}\ }\textbf {\bibinfo {volume} {29}},\ \bibinfo
  {pages} {4260} (\bibinfo {year} {1984})}\BibitemShut {NoStop}%
\bibitem [{\citenamefont {Saha}(1949)}]{Saha_PR_1949}%
  \BibitemOpen
  \bibfield  {author} {\bibinfo {author} {\bibfnamefont {M.~N.}\ \bibnamefont
  {Saha}},\ }\href {\doibase 10.1103/PhysRev.75.1968} {\bibfield  {journal}
  {\bibinfo  {journal} {Phys. Rev.}\ }\textbf {\bibinfo {volume} {75}},\
  \bibinfo {pages} {1968} (\bibinfo {year} {1949})}\BibitemShut {NoStop}%
\bibitem [{\citenamefont {Metlitski}\ \emph {et~al.}(2013)\citenamefont
  {Metlitski}, \citenamefont {Kane},\ and\ \citenamefont
  {Fisher}}]{Metlitski_PRB_2013}%
  \BibitemOpen
  \bibfield  {author} {\bibinfo {author} {\bibfnamefont {M.~A.}\ \bibnamefont
  {Metlitski}}, \bibinfo {author} {\bibfnamefont {C.~L.}\ \bibnamefont {Kane}},
  \ and\ \bibinfo {author} {\bibfnamefont {M.~P.~A.}\ \bibnamefont {Fisher}},\
  }\href {\doibase 10.1103/PhysRevB.88.035131} {\bibfield  {journal} {\bibinfo
  {journal} {Phys. Rev. B}\ }\textbf {\bibinfo {volume} {88}},\ \bibinfo
  {pages} {035131} (\bibinfo {year} {2013})}\BibitemShut {NoStop}%
\bibitem [{\citenamefont {Wang}\ \emph {et~al.}(2014)\citenamefont {Wang},
  \citenamefont {Potter},\ and\ \citenamefont {Senthil}}]{Wang_Science_2014}%
  \BibitemOpen
  \bibfield  {author} {\bibinfo {author} {\bibfnamefont {C.}~\bibnamefont
  {Wang}}, \bibinfo {author} {\bibfnamefont {A.~C.}\ \bibnamefont {Potter}}, \
  and\ \bibinfo {author} {\bibfnamefont {T.}~\bibnamefont {Senthil}},\
  }\href@noop {} {\bibfield  {journal} {\bibinfo  {journal} {Science}\ }\textbf
  {\bibinfo {volume} {343}},\ \bibinfo {pages} {629} (\bibinfo {year}
  {2014})}\BibitemShut {NoStop}%
\bibitem [{\citenamefont {Metlitski}\ and\ \citenamefont
  {Vishwanath}(2016)}]{Metlitski_PRB_2016}%
  \BibitemOpen
  \bibfield  {author} {\bibinfo {author} {\bibfnamefont {M.~A.}\ \bibnamefont
  {Metlitski}}\ and\ \bibinfo {author} {\bibfnamefont {A.}~\bibnamefont
  {Vishwanath}},\ }\href {\doibase 10.1103/PhysRevB.93.245151} {\bibfield
  {journal} {\bibinfo  {journal} {Phys. Rev. B}\ }\textbf {\bibinfo {volume}
  {93}},\ \bibinfo {pages} {245151} (\bibinfo {year} {2016})}\BibitemShut
  {NoStop}%
\bibitem [{\citenamefont {Maciejko}\ and\ \citenamefont
  {Fiete}(2015)}]{Maciejko_NP_2015}%
  \BibitemOpen
  \bibfield  {author} {\bibinfo {author} {\bibfnamefont {J.}~\bibnamefont
  {Maciejko}}\ and\ \bibinfo {author} {\bibfnamefont {G.~A.}\ \bibnamefont
  {Fiete}},\ }\href@noop {} {\bibfield  {journal} {\bibinfo  {journal} {Nature
  Physics}\ }\textbf {\bibinfo {volume} {11}},\ \bibinfo {pages} {385}
  (\bibinfo {year} {2015})}\BibitemShut {NoStop}%
\bibitem [{\citenamefont {Maciejko}\ \emph {et~al.}(2010)\citenamefont
  {Maciejko}, \citenamefont {Qi}, \citenamefont {Karch},\ and\ \citenamefont
  {Zhang}}]{Maciejko_PRL_2010}%
  \BibitemOpen
  \bibfield  {author} {\bibinfo {author} {\bibfnamefont {J.}~\bibnamefont
  {Maciejko}}, \bibinfo {author} {\bibfnamefont {X.-L.}\ \bibnamefont {Qi}},
  \bibinfo {author} {\bibfnamefont {A.}~\bibnamefont {Karch}}, \ and\ \bibinfo
  {author} {\bibfnamefont {S.-C.}\ \bibnamefont {Zhang}},\ }\href {\doibase
  10.1103/PhysRevLett.105.246809} {\bibfield  {journal} {\bibinfo  {journal}
  {Phys. Rev. Lett.}\ }\textbf {\bibinfo {volume} {105}},\ \bibinfo {pages}
  {246809} (\bibinfo {year} {2010})}\BibitemShut {NoStop}%
\bibitem [{\citenamefont {Swingle}\ \emph {et~al.}(2011)\citenamefont
  {Swingle}, \citenamefont {Barkeshli}, \citenamefont {McGreevy},\ and\
  \citenamefont {Senthil}}]{Swingle_PRB_2011}%
  \BibitemOpen
  \bibfield  {author} {\bibinfo {author} {\bibfnamefont {B.}~\bibnamefont
  {Swingle}}, \bibinfo {author} {\bibfnamefont {M.}~\bibnamefont {Barkeshli}},
  \bibinfo {author} {\bibfnamefont {J.}~\bibnamefont {McGreevy}}, \ and\
  \bibinfo {author} {\bibfnamefont {T.}~\bibnamefont {Senthil}},\ }\href
  {\doibase 10.1103/PhysRevB.83.195139} {\bibfield  {journal} {\bibinfo
  {journal} {Phys. Rev. B}\ }\textbf {\bibinfo {volume} {83}},\ \bibinfo
  {pages} {195139} (\bibinfo {year} {2011})}\BibitemShut {NoStop}%
\bibitem [{\citenamefont {Bhattacharjee}\ \emph {et~al.}(2012)\citenamefont
  {Bhattacharjee}, \citenamefont {Kim}, \citenamefont {Lee},\ and\
  \citenamefont {Lee}}]{Bhattacharjee_PRB_2012}%
  \BibitemOpen
  \bibfield  {author} {\bibinfo {author} {\bibfnamefont {S.}~\bibnamefont
  {Bhattacharjee}}, \bibinfo {author} {\bibfnamefont {Y.~B.}\ \bibnamefont
  {Kim}}, \bibinfo {author} {\bibfnamefont {S.-S.}\ \bibnamefont {Lee}}, \ and\
  \bibinfo {author} {\bibfnamefont {D.-H.}\ \bibnamefont {Lee}},\ }\href
  {\doibase 10.1103/PhysRevB.85.224428} {\bibfield  {journal} {\bibinfo
  {journal} {Phys. Rev. B}\ }\textbf {\bibinfo {volume} {85}},\ \bibinfo
  {pages} {224428} (\bibinfo {year} {2012})}\BibitemShut {NoStop}%
\bibitem [{\citenamefont {Huckestein}(1995)}]{Huckestein_RMP_1995}%
  \BibitemOpen
  \bibfield  {author} {\bibinfo {author} {\bibfnamefont {B.}~\bibnamefont
  {Huckestein}},\ }\href {\doibase 10.1103/RevModPhys.67.357} {\bibfield
  {journal} {\bibinfo  {journal} {Rev. Mod. Phys.}\ }\textbf {\bibinfo {volume}
  {67}},\ \bibinfo {pages} {357} (\bibinfo {year} {1995})}\BibitemShut
  {NoStop}%
\bibitem [{\citenamefont {Prange}(1987)}]{Prange_1987}%
  \BibitemOpen
  \bibfield  {author} {\bibinfo {author} {\bibfnamefont {R.~E.}\ \bibnamefont
  {Prange}},\ }in\ \href@noop {} {\emph {\bibinfo {booktitle} {The Quantum Hall
  Effect}}}\ (\bibinfo  {publisher} {Springer},\ \bibinfo {year} {1987})\ pp.\
  \bibinfo {pages} {69--99}\BibitemShut {NoStop}%
\bibitem [{\citenamefont {Ryu}\ \emph {et~al.}(2010)\citenamefont {Ryu},
  \citenamefont {Schnyder}, \citenamefont {Furusaki},\ and\ \citenamefont
  {Ludwig}}]{Ryu_NJP_2010}%
  \BibitemOpen
  \bibfield  {author} {\bibinfo {author} {\bibfnamefont {S.}~\bibnamefont
  {Ryu}}, \bibinfo {author} {\bibfnamefont {A.~P.}\ \bibnamefont {Schnyder}},
  \bibinfo {author} {\bibfnamefont {A.}~\bibnamefont {Furusaki}}, \ and\
  \bibinfo {author} {\bibfnamefont {A.~W.~W.}\ \bibnamefont {Ludwig}},\ }\href
  {http://stacks.iop.org/1367-2630/12/i=6/a=065010} {\bibfield  {journal}
  {\bibinfo  {journal} {New Journal of Physics}\ }\textbf {\bibinfo {volume}
  {12}},\ \bibinfo {pages} {065010} (\bibinfo {year} {2010})}\BibitemShut
  {NoStop}%
\bibitem [{\citenamefont {Agarwala}\ \emph {et~al.}(2017)\citenamefont
  {Agarwala}, \citenamefont {Haldar},\ and\ \citenamefont
  {Shenoy}}]{Agarwala_AOP_2017}%
  \BibitemOpen
  \bibfield  {author} {\bibinfo {author} {\bibfnamefont {A.}~\bibnamefont
  {Agarwala}}, \bibinfo {author} {\bibfnamefont {A.}~\bibnamefont {Haldar}}, \
  and\ \bibinfo {author} {\bibfnamefont {V.~B.}\ \bibnamefont {Shenoy}},\
  }\href {\doibase https://doi.org/10.1016/j.aop.2017.07.016} {\bibfield
  {journal} {\bibinfo  {journal} {Annals of Physics}\ }\textbf {\bibinfo
  {volume} {385}},\ \bibinfo {pages} {469 } (\bibinfo {year}
  {2017})}\BibitemShut {NoStop}%
\end{thebibliography}%

\end{document}